\documentclass[10pt,twocolumn,letterpaper]{article}

\usepackage[pagenumbers]{cvpr} %

\newcommand{\modelname}{PRISM}

\definecolor{cvprblue}{rgb}{0.21,0.49,0.74}
\usepackage[pagebackref,breaklinks,colorlinks,allcolors=cvprblue]{hyperref}

\def\confName{CVPR}
\def\confYear{2027}

\title{PRISM: Perceptual Rate-Controllable Extreme Image Compression \\ with Visual State-Space Models}

\author{Md Adnan Faisal Hossain \qquad Fengqing Zhu\\
Elmore Family School of Electrical and Computer Engineering\\
Purdue University\\
{\tt\small \{hossai34, zhu0\}@purdue.edu}
}

\begin{document}
\maketitle
\begin{abstract}

Generative codecs built on pretrained diffusion priors achieve high perceptual quality at ultra-low bitrates (below 0.05\,bpp). However, most methods still require a separate model for each bitrate, and offer no control over the amount of detail synthesized. Their rate efficiency also depends on the entropy model, where Mamba-based state-space models (SSMs) have recently improved autoregressive context modeling. However, conventional SSM scans are not aligned with the decoding order, and each decoding pass restarts from an empty state, discarding context accumulated in previous passes. We propose \textbf{PRISM}, an extreme image compression method that addresses these limitations. A single model provides two continuous controls: a rate knob that traverses the rate–distortion curve and a realism knob that adjusts the level of detail synthesized. Its visual state-space (VSS) context model introduces a decode-order-aligned (DOA) scan, which predicts each latent from a state that summarizes all decoded context, and cross-tap state carry, which initializes each tap with a summary of the previous one. Experiments show that PRISM outperforms both variable-rate and fixed-rate codecs, requiring on average $22\%$ less bitrate than the strongest baseline at equal FID while also achieving higher PSNR.

\vspace{-0.5cm}

\end{abstract}

\section{Introduction}
\label{sec:intro}
\vspace{-0.1cm}

\begin{figure*}[t]
  \centering
  \includegraphics[width=0.75\linewidth]{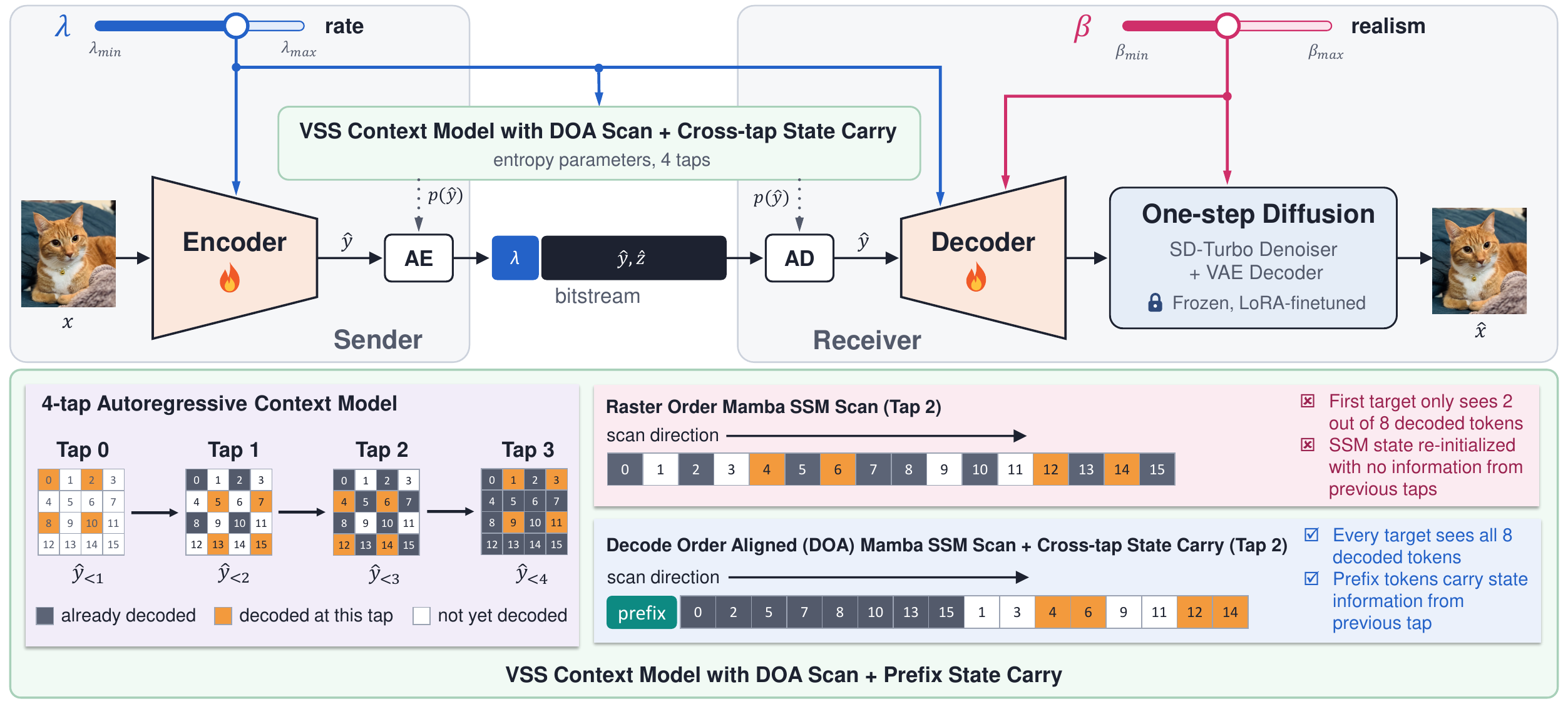}
  \caption{\textbf{Overview of PRISM.}
    \emph{Top:} The sender encodes $x$ under a continuous rate control $\lambda$.
    An autoregressive VSS context model predicts the entropy parameters of $\hat{y}$.
    The receiver reconstructs $\hat{x}$ with a one-step diffusion decoder. A realism control $\beta$ sets how much detail is synthesized.
    \emph{Bottom:} At tap~2, the forward raster scan lets the first target see only 2 of the 8 decoded tokens, and the SSM state restarts at every tap.
    The DOA scan visits all decoded tokens before any target, and prefix tokens carry state from the previous tap.}
  \label{fig:overview}
  \vspace{-0.4cm}
\end{figure*}

Over the past few years, learned image compression~\cite{balle2018variational, minnen2018joint, cheng2020learned, he2022elic} has surpassed traditional codecs such as JPEG~\cite{wallace1991jpeg}, BPG\footnote{\url{https://bellard.org/bpg/}}, and VTM~\cite{bross2021overview} in rate--distortion performance. At extremely low bitrates, however, distortion-optimized codecs produce blurry reconstructions, since the MSE-optimal output is the average of all plausible images. Generative codecs avoid this by also matching reconstructions to the distribution of natural images, trading a small increase in distortion for much higher realism~\cite{blau2019rethinking}. Although early generative codecs relied on GANs~\cite{agustsson2019generative, mentzer2020high, muckley2023improving}, more recent ones build on pretrained diffusion models~\cite{rombach2022high}. These models can reach high perceptual quality at ultra-low bitrates ($\leq 0.05$\,bpp)~\cite{zhang2025stablecodec}, with some even achieving single-step decoding~\cite{zhang2025stablecodec, guo2025oscar, zhang2026ultra}.

Despite this progress, two limitations remain. First, these codecs offer little control at deployment. Most train a separate model for each target bitrate~\cite{zhang2025stablecodec, zhang2026ultra}, multiplying storage and training costs. The few variable-rate generative codecs support only a handful of discrete rates~\cite{cao2026progic}, need per-rate sub-modules~\cite{guo2025oscar}, or decode with many denoising steps~\cite{vonderfecht2025lossy}, and all of them trail the best fixed-rate models in coding performance. Most methods also fix the balance between fidelity and realism at training time, so users cannot choose how much detail is synthesized.

Second, the rate also depends on the entropy model, whose context model decodes the latent autoregressively in a few parallel passes (\emph{taps}). Mamba state-space models (SSMs)~\cite{gu2023mamba} have recently strengthened these context models~\cite{zeng2025mambaic, chen2026content} by summarizing the scanned sequence in a hidden state at linear cost. However, they typically scan in raster order, which does not match the decoding order: the state reaching an undecoded latent holds only the decoded latents that happen to precede it in the scan, so no single state holds all of its decoded context (\cref{fig:overview}, bottom). Moreover, each tap restarts its scan from an empty state, discarding the context summarized at the previous tap.

We propose PRISM, a perceptual, rate-controllable extreme image codec with a one-step diffusion decoder that addresses both limitations (\cref{fig:overview}). A single model covers a continuous range of bitrates using a rate knob $\lambda$ that conditions the encoder, decoder and the entropy model, and sets a per-channel quantization step $\Delta_\lambda$.  Because $\lambda$ is sampled from a continuous range during training, the model supports any rate within that range rather than a few discrete points, without per-rate models or sub-modules. PRISM also exposes a realism knob $\beta$ that conditions only the decoder, so the same bitstream can be decoded anywhere between a fidelity-oriented and a realism-oriented reconstruction.

To address the shortcomings of existing SSM context models, PRISM also introduces a visual state-space (VSS) context model with two lightweight components: a decode-order-aligned (DOA) scan and cross-tap state carry. The DOA scan ensures that all previously decoded latents precede the target in the scan order, so that it is predicted from a state that summarizes all decoded context. To extend context across taps, cross-tap state carry summarizes the previous tap's output into a small set of tokens and prepends them to the next tap's scan. This allows each tap to build upon the context accumulated by its predecessor, analogous to hidden-state propagation in recurrent networks. 
Our contributions can be summarized as follows:
\begin{itemize}
    \item We propose PRISM, a one-step diffusion codec in which a single model offers continuous control over rate and perceptual quality. Rate is controlled through $\lambda$-conditioned transforms and quantization steps, while perceptual quality is adjusted through a decoder-side parameter $\beta$.
    \item We introduce a visual state-space (VSS) context model comprising a decode-order-aligned (DOA) scan and cross-tap state carry. The DOA scan ensures that each latent is predicted from a state that summarizes all decoded context, while cross-tap state carry propagates accumulated context across successive taps through a compact hidden-state representation.
    \item Experiments on three benchmark datasets show that PRISM outperforms existing variable-rate and fixed-rate codecs across several perceptual metrics. 
\end{itemize}

\section{Related Work}
\label{sec:related_work}
\vspace{-0.1cm}

\begin{figure*}[t]
  \centering
  \includegraphics[width=0.9\linewidth]{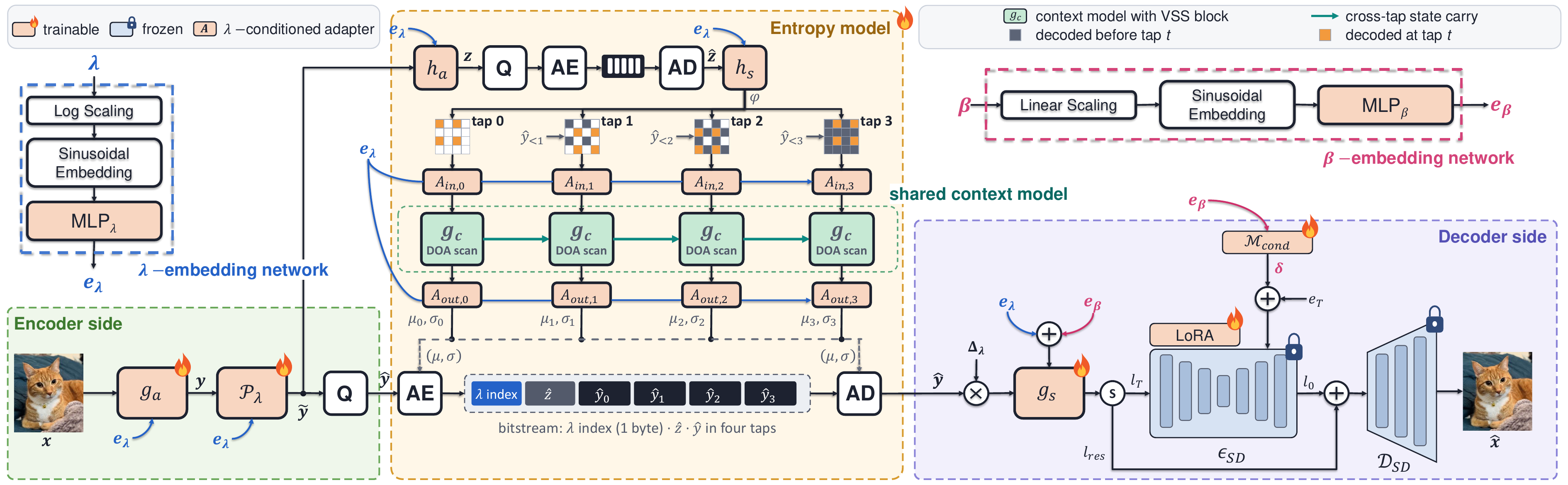}
  \caption{\textbf{Network architecture of \modelname{}.}
    The rate control $\lambda$ is embedded as $\mathbf{e}_\lambda$.
    It conditions the analysis transform $g_a$, the rate-adaptive scaling $\mathcal{P}_\lambda$, the hyper transforms $h_a$ and $h_s$, the shared context model $g_c$ and the synthesis transform $g_s$. The realism control $\beta$ is embedded as $\mathbf{e}_\beta$ and conditions only the decoder side, $g_s$ and, through $\mathcal{M}_{\mathrm{cond}}$, the denoiser $\epsilon_{\mathrm{SD}}$. $g_c$ runs in four taps with DOA scans, cross-tap state carry and per-tap $\lambda$-conditioned adapters $A_{\mathrm{in},t}$ and $A_{\mathrm{out},t}$.
    The bitstream holds a one-byte $\lambda$ index, $\hat{z}$, and $\hat{y}$.}
  \label{fig:architecture}
  \vspace{-0.3cm}
\end{figure*}

\subsection{Generative Image Compression}
\label{sec:rel_work_generative_compression}
\vspace{-0.1cm}

At a fixed rate, reconstructions can be made more realistic only by accepting higher distortion~\cite{blau2018perception, blau2019rethinking}. Generative codecs make this trade with GAN~\cite{santurkar2018generative, agustsson2019generative, mentzer2020high, he2022po, muckley2023improving, korber2024egic} or VQ~\cite{jia2024generative, xue2025dlf} decoders and, more recently, with diffusion decoders, trained from scratch~\cite{theis2022lossy, yang2023lossy, hoogeboom2023high, jia2026cod} or built on a pretrained text-to-image model~\cite{rombach2022high}, often with text or semantic side information~\cite{lei2023text, careil2024towards, korber2024perco, li2024toward, relic2024lossy, xia2025diffpc, ke2025ultra, xu2025picd, sheng2026cadc}. Training-free methods~\cite{xu2024idempotence, vonderfecht2025lossy, ohayon2025compressed, vaisman2026turbo} need many denoising steps, whereas one-step codecs~\cite{zhang2025stablecodec, xue2025one, guo2025oscar, chen2026steering, park2026single, zhang2026ultra} fine-tune the denoiser to decode in a single pass. \citet{yang2026representation} minimize the Fr\'echet distance in a representation space directly; we adapt this loss to the compression task.

\subsection{Rate and Realism Control}
\label{sec:rel_work_var_rate}
\vspace{-0.1cm}

Distortion-oriented codecs reach variable rates by conditioning on $\lambda$~\cite{choi2019variable, song2021variable, duan2023qarv} or by interpolating learned per-channel gains~\cite{cui2021asymmetric}. Generative codecs are less flexible: GAN and VQ codecs offer discrete quality levels~\cite{iwai2024controlling, cao2026progic} or adapt the granularity of their codes~\cite{li2025once}, and diffusion codecs are trained at a few rates~\cite{relic2024lossy, chang2025generative, xu2025decouple}, need more denoising steps at lower rates~\cite{relic2025bridging}, or train a hyper-encoder per rate~\cite{guo2025oscar}. Training-free codecs vary the rate freely but trail trained codecs in quality~\cite{vonderfecht2025lossy, ohayon2025compressed, vaisman2026turbo}. Since one representation can serve a range of distortion--realism trade-offs~\cite{zhang2021universal}, realism can also be set at decoding time: GAN codecs~\cite{iwai2021fidelity, agustsson2023multi, iwai2024controlling} do so, MRIC~\cite{agustsson2023multi} through a realism weight given to the decoder, and \citet{zhou2025controllable} blend in the features of a diffusion module. \modelname{} controls both rate and realism continuously in a one-step diffusion codec.

\subsection{Entropy Models}
\label{sec:rel_work_entropy_models}
\vspace{-0.1cm}

Autoregressive context models~\cite{minnen2018joint, cheng2020learned} condition each latent on the hyperprior~\cite{balle2018variational} and on the latents already decoded, and parallel variants decode in a few passes over spatial~\cite{he2021checkerboard}, channel~\cite{minnen2020channel} or space--channel groups~\cite{he2022elic, lu2022high, jiang2023mlic}. AEIC~\cite{zhang2026ultra}, for example, uses four space--channel taps with a convolutional context model applied afresh at every tap. For longer-range context, codecs use attention~\cite{jiang2023mlic, feng2025linear} or state-space models (SSMs)~\cite{gu2023mamba, liu2024vmamba}. SSMs have been used in the transforms~\cite{qin2024mambavc, qin2025cassic} and context models~\cite{zeng2025mambaic, chen2026content}, and SSM or RWKV blocks in lightweight generative codecs~\cite{wei2026lightweight, zhang2026samic, liu2026mrt}. These context models scan in fixed~\cite{zeng2025mambaic} or content-adaptive~\cite{chen2026content} orders rather than the decoding order, and start every pass from an empty state. \modelname{} aligns each scan with the decoding order and carries the state across taps.

\section{Background}
\label{sec:background}
\vspace{-0.1cm}

\subsection{Image Compression with Diffusion Priors}
\label{sec:background_diffusion_codec}
\vspace{-0.1cm}

Latent diffusion models~\cite{rombach2022high} such as Stable Diffusion (SD) work in the latent space of a VAE, whose encoder maps an image $x$ to a latent $l_0$ at $1/8$ of its resolution and whose decoder $\mathcal{D}_{\mathrm{SD}}$ maps it back. The forward process gradually replaces the latent with Gaussian noise,
\begin{equation}
  l_t = \sqrt{\bar{\alpha}_t}\, l_0 + \sqrt{1 - \bar{\alpha}_t}\, \epsilon, \quad \epsilon \sim \mathcal{N}(0, I),
  \label{eq:diffusion_forward}
\end{equation}
where the noise schedule $\bar{\alpha}_t$ falls from $1$ to almost $0$ at $t = T$, and a denoiser $\epsilon_\theta$ is trained to predict $\epsilon$ from $l_t$ and $t$~\cite{ho2020denoising}. Each prediction gives an estimate of the clean latent,
\begin{equation}
  \hat{l}_0 = \bigl(l_t - \sqrt{1 - \bar{\alpha}_t}\, \epsilon_\theta(l_t, t)\bigr) / \sqrt{\bar{\alpha}_t},
  \label{eq:diffusion_x0}
\end{equation}
which a sampler refines over many steps, starting from pure noise at $t = T$. SD-Turbo~\cite{sauer2024adversarial} is distilled so that a single step from $t = T$ already gives a sharp $\hat{l}_0$. One-step codecs~\cite{zhang2025stablecodec, guo2025oscar, zhang2026ultra} feed the denoiser a latent decoded from the bitstream in place of noise and fine-tune it to map this latent to $l_0$ in a single pass. Since the bitstream already carries the structure of the image, the denoiser only has to synthesize, from its diffusion prior, the details the bitstream leaves out.

\subsection{Mamba State-Space Models}
\label{sec:background_mamba}
\vspace{-0.1cm}

A state-space model (SSM)~\cite{gu2022efficiently} maps an input $u(t) \in \mathbb{R}$ to an output $o(t) \in \mathbb{R}$ through a hidden state $h(t) \in \mathbb{R}^{n}$,
\begin{equation}
  h'(t) = A\, h(t) + B\, u(t), \quad o(t) = C\, h(t) + D\, u(t),
  \label{eq:ssm_continuous}
\end{equation}
with $A \in \mathbb{R}^{n \times n}$, $B \in \mathbb{R}^{n \times 1}$, $C \in \mathbb{R}^{1 \times n}$ and $D \in \mathbb{R}$. For a sequence $u_1, \dots, u_L$ sampled with step $\Delta$, the zero-order hold, which assumes $u$ constant over each step, discretizes $A$ and $B$ into $\bar{A} = \exp(\Delta A)$ and $\bar{B} = (\Delta A)^{-1}(\exp(\Delta A) - I)\, \Delta B$, which in practice is simplified to $\Delta B$. Mamba~\cite{gu2023mamba} makes the SSM selective by computing $\Delta_j$, $B_j$ and $C_j$ from each token $u_j$, which gives the recurrence
\begin{align}
  h_j &= \exp(\Delta_j A)\, h_{j-1} + \Delta_j B_j\, u_j, \label{eq:ssm_state} \\
  o_j &= C_j\, h_j + D\, u_j, \label{eq:ssm_output}
\end{align}
with $h_0 = 0$ and a learned diagonal $A$. Each token thus decides what its state keeps and what it forgets, while the scan, run independently for each channel, stays linear in the number of tokens $L$. VMamba~\cite{liu2024vmamba} extends Mamba to images with the VSS block, whose 2D selective scan flattens a feature map along four routes, row- and column-major in both directions, scans each of them and merges their outputs.
\section{Method}
\label{sec:method}
\vspace{-0.1cm}

This section describes \modelname{}. \Cref{sec:method_overview} gives an overview of the codec and its networks. \Cref{sec:method_rate,sec:method_realism} explain the rate and realism control mechanisms respectively. \cref{sec:method_context_model} introduces the proposed context model with the DOA scan and the cross-tap state carry, and \cref{sec:method_training_objective} describes our training objective.

\subsection{Overview}
\label{sec:method_overview}
\vspace{-0.1cm}

\begin{figure}[t]
  \centering
  \includegraphics[width=0.9\linewidth]{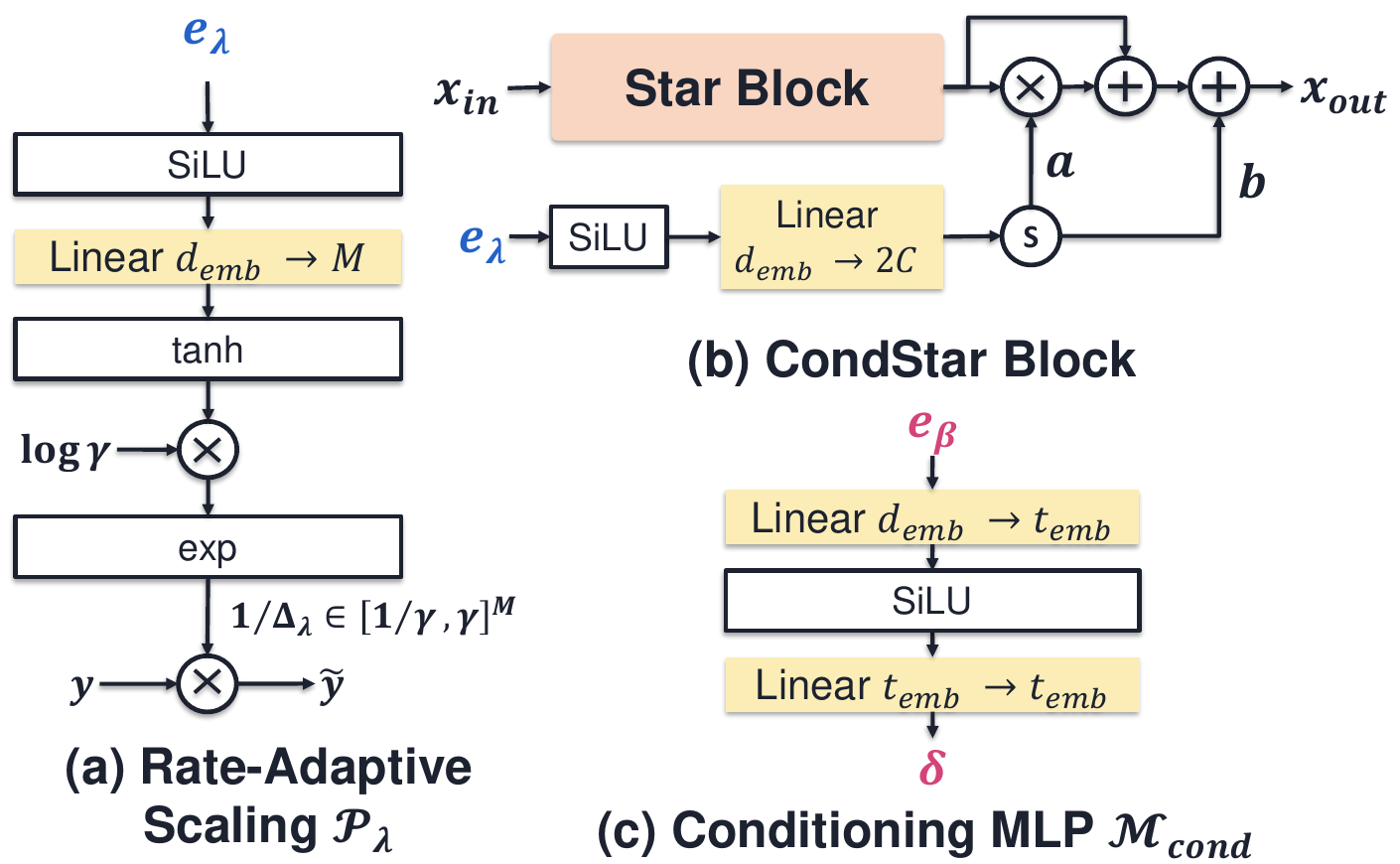}
  \vspace{-0.1cm}
  \caption{\textbf{Conditioning modules.} (a)~Rate-adaptive scaling predicts the inverse quantization step $\Delta_\lambda^{-1} \in [1/\gamma, \gamma]^M$ from $\mathbf{e}_\lambda$. (b)~A CondStar block modulates the output of a Star block with a scale and a shift predicted from $\mathbf{e}_\lambda$. (c)~The conditioning MLP maps $\mathbf{e}_\beta$ to the offset $\delta$ on the timestep embedding of the denoiser.}
  \label{fig:film_ras}
  \vspace{-0.3cm}
\end{figure}

\Cref{fig:overview} summarizes \modelname{}, and \cref{fig:architecture} shows its network architecture. A single model codes an image at any rate in a continuous range, set by the rate control $\lambda$ at the sender, and decodes it at any level of realism, set by the realism control $\beta$ at the receiver.

\vspace{-0.4cm}
\paragraph{Encoding-Decoding}
The analysis transform $g_a$ maps the image $x \in \mathbb{R}^{3 \times H \times W}$ to an $M$-channel latent $y \in \mathbb{R}^{M \times H/32 \times W/32}$. The rate-adaptive scaling block $\mathcal{P}_\lambda$ then applies a learned per-channel step $\Delta_\lambda$ to the latent to form the scaled latent $\tilde{y} = y \oslash \Delta_\lambda$, which is quantized to $\hat{y}$ and entropy coded. The hyper analysis transform $h_a$ maps $\tilde{y}$ to a quantized side latent $\hat{z}$, which is entropy coded with a factorized prior $p_{\hat{z}}(\hat{z})$~\cite{balle2018variational, minnen2018joint}, and the hyper synthesis transform $h_s$ turns $\hat{z}$ into features $\varphi$. The latent $\hat{y}$ is coded autoregressively in four taps,
\vspace{-0.2cm}
\begin{equation}
\vspace{-0.2cm}
  p_{\hat{y}|\hat{z}}(\hat{y} \mid \hat{z}) = \prod_{t=0}^{3} p\bigl(\hat{y}_t \mid \hat{y}_{<t}, \hat{z}\bigr),
  \label{eq:taps}
\end{equation}
where $\hat{y}_t$ denotes the elements coded at tap $t$. Each element is modeled as a Gaussian $\mathcal{N}(\mu, \sigma^2)$ convolved with a unit-width uniform distribution~\cite{balle2018variational}, whose mean and scale are predicted by the context model from $\varphi$ and the elements of the previous taps, $(\mu_t, \sigma_t) = g_c(\varphi, \hat{y}_{<t}; \lambda)$. The rate-control knob is also quantized and sent as a one-byte index in the bitstream. The receiver reads $\lambda$ and decodes $\hat{y}$ with the same autoregressive entropy model. It rescales $\hat{y}$ by $\Delta_\lambda$, and the synthesis transform $g_s$ maps it to the input $l_T$ of the denoiser and a residual $l_{\mathrm{res}}$. The denoiser $\epsilon_{\mathrm{SD}}$ is the UNet of SD-Turbo~\cite{sauer2024adversarial}, fine-tuned with LoRA~\cite{hu2022lora}, and it maps $l_T$ to the clean latent $l_0$ in a single step. A frozen VAE decoder $\mathcal{D}_{\mathrm{SD}}$ from~\citet{chen2025adversarial} then decodes $l_0 + l_{\mathrm{res}}$ into the reconstruction $\hat{x}$.

\begin{figure}[t]
  \centering
  \includegraphics[width=0.7\linewidth]{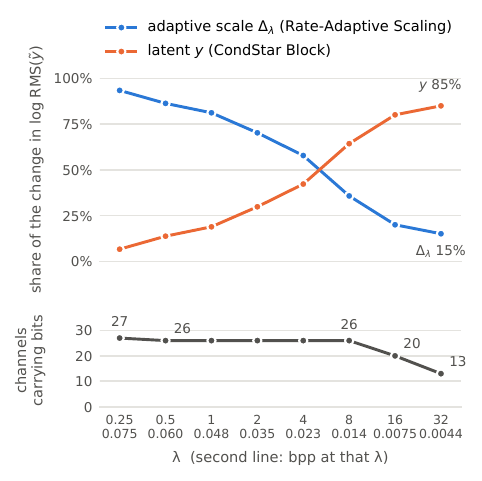}
  \vspace{-0.3cm}
  \caption{\textbf{Two rate controls} on Kodak. \emph{Top:} Share of the change in the scaled latent $\tilde{y}$ between consecutive rate-points due to each mechanism. \emph{Bottom:} Active channels (more than $1\%$ of elements of $\tilde{y}$ round to nonzero) at each rate-point.}
  \vspace{-0.4cm}
  \label{fig:gain_share}
\end{figure}

\vspace{-0.4cm}
\paragraph{Rate and Realism Controls}
We map the scalar knobs $\lambda$ and $\beta$ into higher-dimensional embeddings $\mathbf{e}_\lambda$ and $\mathbf{e}_\beta$ using logarithmic and linear scaling, respectively, followed by sinusoidal embeddings and MLPs. Unlike scalar conditioning, which tends to capture only slowly varying dependencies, sinusoidal embeddings enable sensitivity to fine-grained changes~\cite{tancik2020fourier,ho2020denoising}. $\mathbf{e}_\lambda$ conditions every network that shapes the bitstream, $g_a$, $\mathcal{P}_\lambda$, $h_a$, $h_s$, the entropy model and $g_s$, while $\mathbf{e}_\beta$ conditions only the decoder side, $g_s$ and $\epsilon_{\mathrm{SD}}$.

\begin{figure*}[t]
  \centering
  \includegraphics[width=0.85\linewidth]{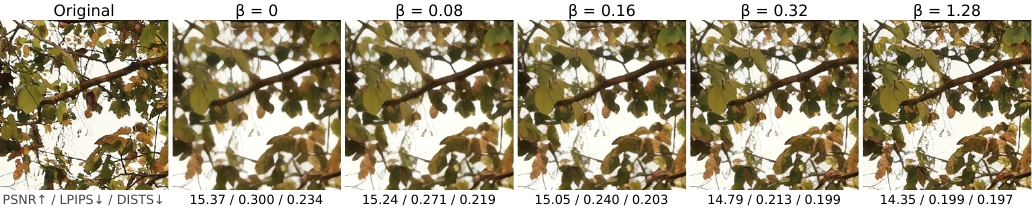}
  \vspace{-0.2cm}
  \caption{\textbf{One bitstream, decoded at five values of $\beta$.} Results obtained using a $256 \times 256$ crop of a DIV2K image coded at $\lambda = 0.25$ ($0.0905$\,bpp). The bitstream is identical for all reconstructions; only the decoder's realism control $\beta$ changes.}
  \label{fig:same_bitstream}
  \vspace{-0.3cm}
\end{figure*}

\vspace{-0.4cm}
\paragraph{Context Model}
The context model $g_c$, each of whose blocks contains a Mamba VSS block~\cite{liu2024vmamba}, is shared by the four taps. At tap $t$, a $\lambda$-conditioned adapter $A_{\mathrm{in},t}$ maps $\varphi$ and $\hat{y}_{<t}$ to the input of $g_c$, and a second adapter $A_{\mathrm{out},t}$ maps its output to the means $\mu_t$ and scales $\sigma_t$ of the elements coded at tap $t$. Network details are in \cref{sec:supp_network_details}.

\subsection{Rate Control}
\label{sec:method_rate}
\vspace{-0.1cm}

The rate of the latent is set by its size relative to the quantization step. A codec can therefore lower the rate by enlarging the step~\cite{choi2019variable, cui2021asymmetric} or by shrinking the latent~\cite{choi2019variable, song2021variable, duan2023qarv}. \modelname{} does both: rate-adaptive scaling sets a per-channel step, and CondStar blocks adapt the transforms. 

\vspace{-0.4cm}
\paragraph{Rate-Adaptive Scaling}
$\mathcal{P}_\lambda$ maps the rate embedding $\mathbf{e}_\lambda$ to the inverse of a per-channel quantization step (\cref{fig:film_ras}a):
\begin{equation}
  \Delta_\lambda^{-1} = \exp\bigl(\log \gamma \cdot \tanh\bigl(W\, \mathrm{SiLU}(\mathbf{e}_\lambda) + b\bigr)\bigr),
  \label{eq:ras}
\end{equation}
which lies in $[1/\gamma,\, \gamma]^M$. The latent is scaled to $\tilde{y} = y \oslash \Delta_\lambda$, so that $\operatorname{Var}(\tilde{y}_c) = \operatorname{Var}(y_c) / \Delta_{\lambda,c}^2$ for each channel $c$, and the bound $\gamma = 8$ limits how far $\Delta_\lambda$ can shift this variance. The decoder multiplies $\hat{y}$ by the same $\Delta_\lambda$. The whole chain acts as a uniform quantizer of $y$ with per-channel step $\Delta_\lambda$. Unlike the gain units of \citet{cui2021asymmetric}, which are learned for a few rates and interpolated, $\Delta_\lambda$ is predicted for any $\lambda$ in $[\lambda_{\min}, \lambda_{\max}] = [0.05, 32]$. We initialize $W$ and $b$ to zero, so $\Delta_\lambda = 1$ at the start of training.

\vspace{-0.4cm}
\paragraph{CondStar Block}
To further support rate-conditional coding, especially at the lowest rates, we introduce CondStar blocks. At low rates, the coded residual $\tilde{y} - \mu$ of the information-carrying channels is often much smaller than the quantization step. The rate then falls mainly because channels stop carrying bits, and that requires changing the latent $y$ itself. CondStar blocks therefore condition the transforms $g_a$, $h_a$, $h_s$ and $g_s$ directly by applying FiLM-style~\cite{perez2018film} modulation to the outputs of their intermediate Star blocks $S$~\cite{ma2024rewrite} (\cref{fig:film_ras}b):
\begin{equation}
\vspace{-0.1cm}
  x_{\mathrm{out}} = (1 + \mathbf{a}) \odot S(x_{\mathrm{in}}) + \mathbf{b}, \quad [\mathbf{a}, \mathbf{b}] = W\, \mathrm{SiLU}(\mathbf{e}),
  \label{eq:condstar}
\end{equation}
where $\mathbf{a}$ and $\mathbf{b}$ are the per-channel scale and shift predicted from the embedding $\mathbf{e}$, with $\mathbf{e} = \mathbf{e}_\lambda$ in $g_a$, $h_a$ and $h_s$, and $\mathbf{e} = \mathbf{e}_\lambda + \mathbf{e}_\beta$ in $g_s$. 

The size of the scaled latent per-channel is $\log \mathrm{RMS}(\tilde{y}_c) = \log \mathrm{RMS}(y_c) - \log \Delta_{\lambda,c}$, so its change between two values of $\lambda$ splits exactly into a part due to $\Delta_\lambda$ and a part due to $y$. From~\cref{fig:gain_share}, we can see that at higher rates (above $0.048$\,bpp), $\Delta_\lambda$ accounts for $81$--$93\%$ of the change, while at the lowest rates, the CondStar blocks take over by switching channels off: $26$ channels carry bits at $\lambda = 8$, but only $13$ at $\lambda = 32$.

\subsection{Realism Control}
\label{sec:method_realism}
\vspace{-0.1cm}

The realism control $\beta$ sets how much detail the decoder synthesizes. Since $\beta$ conditions only the decoder, the same bitstream can be decoded at any $\beta$ without re-encoding (\cref{fig:same_bitstream}). $\beta$ reaches the decoder by two routes. First, the CondStar blocks of $g_s$ receive $\mathbf{e}_\lambda + \mathbf{e}_\beta$ (\cref{eq:condstar}), so $\beta$ shapes both $l_T$ and $l_{\mathrm{res}}$. Second, $\beta$ enters the denoiser $\epsilon_{\mathrm{SD}}$ through its timestep pathway, which in diffusion models signals the input's noise level and thus how much detail to generate. As SD-Turbo runs at a fixed timestep $T = 999$, we fold its timestep embedding $\mathbf{e}_T$ into the biases of the UNet's residual blocks. The conditioning MLP $\mathcal{M}_{\mathrm{cond}}$ (\cref{fig:film_ras}c) maps $\mathbf{e}_\beta$ to an offset $\delta$, so each residual block computes as if its timestep embedding were $\mathbf{e}_T + \delta$ (\cref{fig:architecture}). Implementation details and distortion--perception trade-offs are in \cref{sec:supp_cond_unet}.
\subsection{Context Model}
\label{sec:method_context_model}
\vspace{-0.1cm}

\begin{figure}[t]
  \centering
  \includegraphics[width=0.85\linewidth]{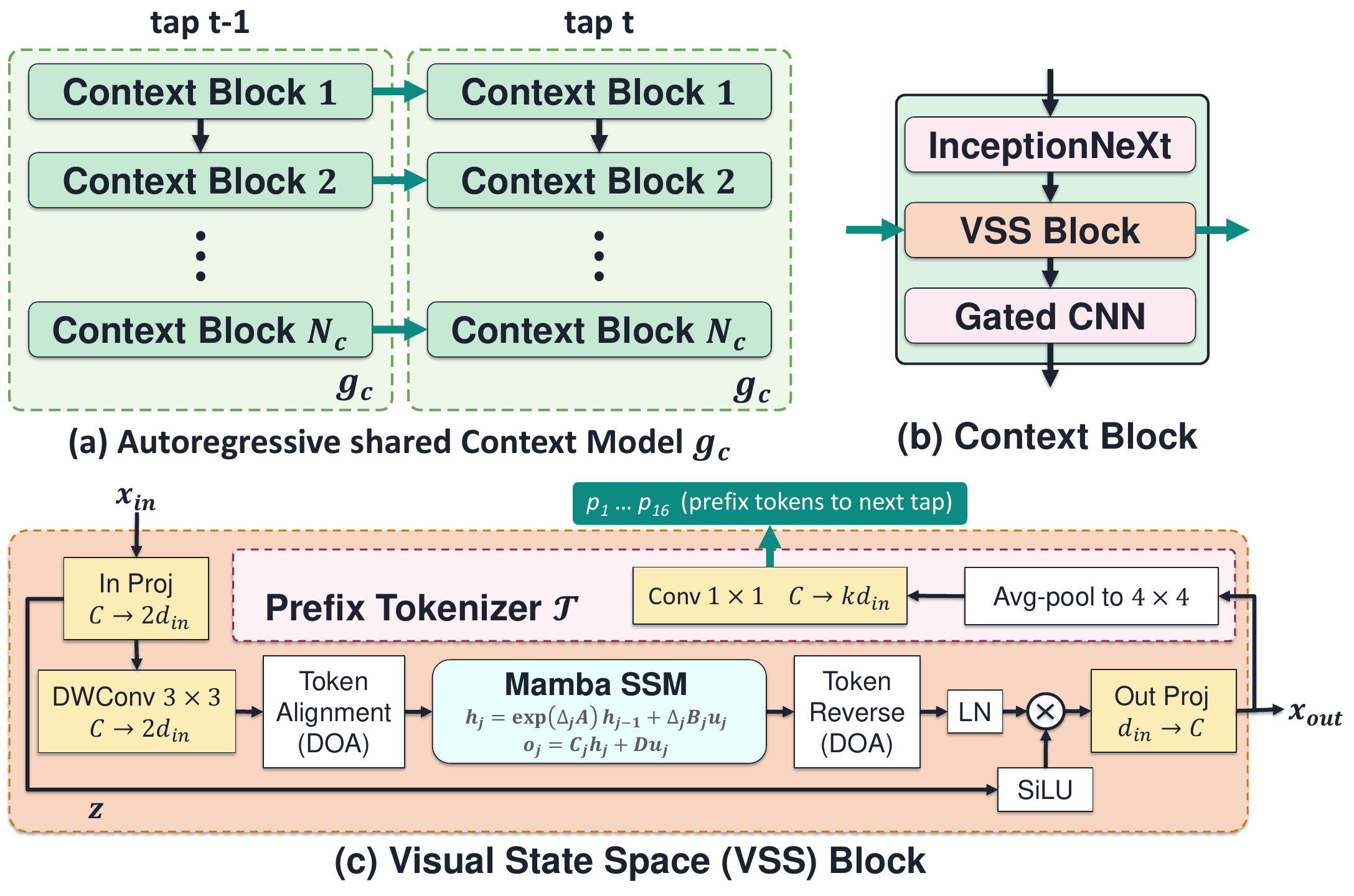}
  \vspace{-0.3cm}
  \caption{\textbf{Context model.} (a)~The context model $g_c$ shared by the four taps, and each of its $N_c$ blocks. (b)~Components of a context block. (c)~The VSS block consisting of a token alignment operation, Mamba SSM layer and a prefix tokenizer.}
  \label{fig:context_model}
  \vspace{-0.4cm}
\end{figure}

\Cref{fig:context_model} shows the context model $g_c$. Each of its $N_c = 4$ blocks mixes information locally with an InceptionNeXt~\cite{yu2024inceptionnext} and a Gated CNN block~\cite{yu2025mambaout}, and globally with a VSS block~\cite{liu2024vmamba} between them. The VSS block flattens its input into spatial tokens $u_1, \dots, u_L$ in some scan order and runs the selective scan of \cref{eq:ssm_state,eq:ssm_output}. The state $h_j$ summarizes the tokens up to position $j$, and what the output $o_j$ of a token can use depends on the scan order.

\vspace{-0.5cm}
\paragraph{Decode-Order-Aligned (DOA) Scan}
In raster order, the state at an undecoded target holds only the decoded elements that precede it: \eg at tap~2 in \cref{fig:overview}, the first target sees only $2$ of the $8$ decoded tokens. The DOA scan reorders the sequence so that all decoded elements come first, and every target is scanned with a state that has absorbed all of them.

The taps, however, interleave channels as well as positions: the latent is split into $G = 4$ channel groups $\hat{y}^1, \dots, \hat{y}^G$, and each tap codes a different set of spatial tokens in each group. Therefore, a single spatial order cannot place the decoded elements first in each group, and DOA gives each channel group its own scan order. Formally, let $\mathcal{K}_{t}^g$ be the set of spatial tokens at which channel group $g$ has been decoded before tap $t$, and let $\mathcal{U}_t^g$ be the remaining ones. The VSS block then scans the features of channel group $g$ in the order $\pi_t^g = \bigl(\mathrm{raster}(\mathcal{K}_t^g), \mathrm{raster}(\mathcal{U}_t^g)\bigr)$, where $\mathrm{raster}(\cdot)$ lists a set of tokens in row-major order. At tap~0, $\pi_0^g$ is plain raster order. As in VMamba, a second direction scans the same order backwards. Unlike raster order, the DOA scan reaches every target with a state that has absorbed all decoded context.

\begin{figure*}[t]
  \centering
  \includegraphics[width=0.95\linewidth]{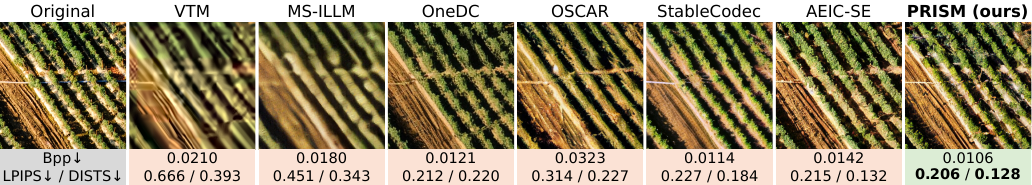}
  \\[2pt]
  \includegraphics[width=0.95\linewidth]{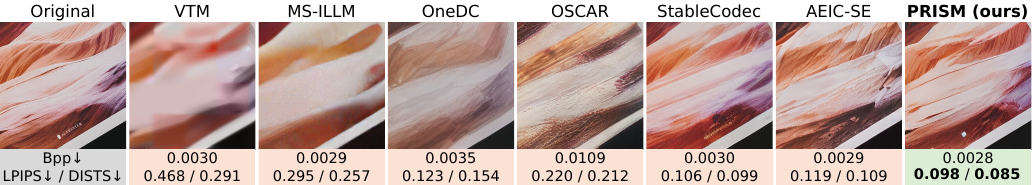}
  \vspace{-0.2cm}
  \caption{\textbf{Qualitative comparison on images from DIV2K (top) and CLIC (bottom).} Images shown are $256 \times 256$ crops. The bpp, LPIPS and DISTS of the whole image are given below each crop, with the best metrics in bold. Each baseline is shown at its lowest rate above that of \modelname{}.}
  \vspace{-0.5cm}
  \label{fig:qual}
\end{figure*}

\vspace{-0.4cm}
\paragraph{Cross-Tap State Carry}
A plain scan restarts its internal state from $h_0 = 0$ at every tap, discarding all previously gathered context. To carry this context over, the prefix tokenizer $\mathcal{T}$ summarizes the block's output at the previous tap, $o^{(t-1)}$, into $L_p = 16$ prefix tokens per direction (\cref{fig:context_model}c):
\vspace{-0.1cm}
\begin{equation}
\vspace{-0.1cm}
  [p_1, \dots, p_{L_p}] = W_p\, \mathrm{AvgPool}_{4 \times 4}\bigl(o^{(t-1)}\bigr),
  \label{eq:prefix}
\end{equation}
where $W_p$ is a $1 \times 1$ convolution. These tokens are prepended to the scan of the current tap $t$. In each channel group $g$, the block scans $(p_1, \dots, p_{L_p}, u_{\pi_t^g(1)}, \dots, u_{\pi_t^g(L)})$ and discards the outputs at the prefix positions. The backward scan starts from its own prefix, and carries the previous tap's summary directly to the targets, bypassing the decoded tokens. Unrolling \cref{eq:ssm_state} over the prefix from $h_0 = 0$ gives the state from which the first spatial token is scanned:
\vspace{-0.2cm}
\begin{equation}
\vspace{-0.1cm}
  h_{L_p} = \sum_{j=1}^{L_p} \Bigl(\prod_{i=j+1}^{L_p} \exp(\Delta_i A)\Bigr) \Delta_j B_j\, p_j.
  \label{eq:prefix_state}
\end{equation}
Each $p_j$ writes $\Delta_j B_j\, p_j$ into the state, which then decays by $\exp(\Delta_i A)$ at each later prefix position $i$. As each $p_j$ summarizes one cell of a $4 \times 4$ grid over $o^{(t-1)}$, $h_{L_p}$ thus summarizes the whole latent at the previous tap.

\subsection{Training Objective}
\label{sec:method_training_objective}
\vspace{-0.1cm}

\begin{figure*}[t]
  \centering
  \includegraphics[width=0.95\linewidth]{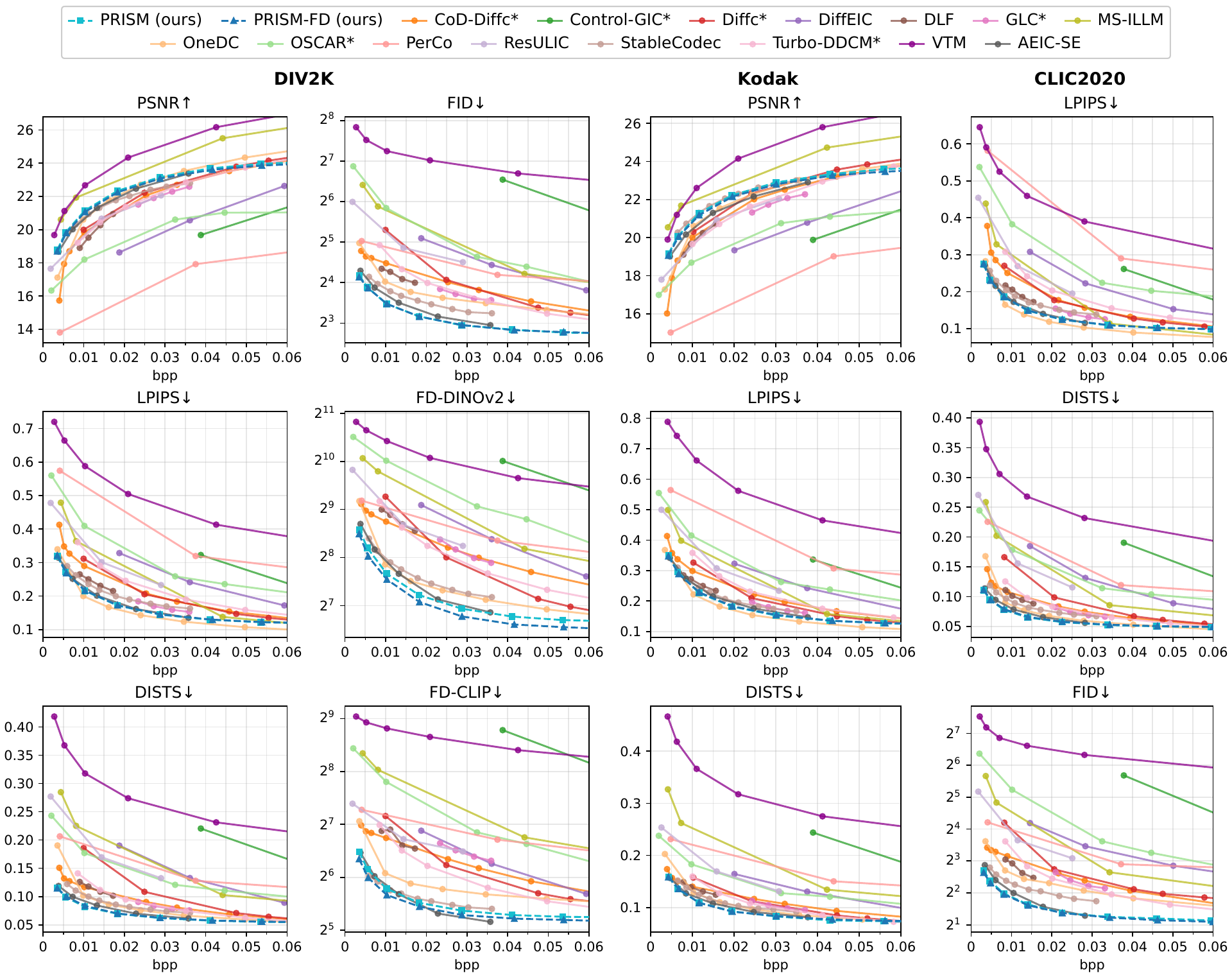}
  \vspace{-0.3cm}
  \caption{\textbf{Rate--distortion--perception comparison on DIV2K (first two columns), Kodak and CLIC2020.}
    $\uparrow$/$\downarrow$ represents higher/lower is better.
    Codecs marked with * are variable-rate; the others, except VTM, train a separate model for each rate.
    \modelname{} and \modelname{}-FD are shown at $\beta = 5.12$ on DIV2K and CLIC2020 and at $\beta = 1.28$ on Kodak.}
    \vspace{-0.5cm}
  \label{fig:rd}
\end{figure*}

\begingroup
\setlength{\abovedisplayskip}{6pt plus 2pt minus 2pt}
\setlength{\belowdisplayskip}{6pt plus 2pt minus 2pt}
\setlength{\abovedisplayshortskip}{0pt plus 2pt}
\setlength{\belowdisplayshortskip}{4pt plus 2pt minus 2pt}
\setlength{\jot}{2pt}

\paragraph{Multi-Stage Training}
We train \modelname{} in three stages: stage~1 pretrains the codec with $\beta=0$, stage~2 adds the realism control, and stage~3 fine-tunes at a higher resolution. The objectives of stages~1 and~2 are:
\begin{align}
  \mathcal{L}_{\mathrm{S1}} &= \lambda\, \mathcal{R}(\hat{y}, \hat{z}) + \mathcal{D}(x, \hat{x}) + \gamma_p\, \mathcal{L}_{p}(x, \hat{x}), \label{eq:loss_s1} \\
  \mathcal{L}_{\mathrm{S2}} &= \lambda\, \mathcal{R}(\hat{y}, \hat{z}) + \mathcal{D}(x, \hat{x}) + \beta\, \mathcal{L}_{\mathrm{real}}, \label{eq:loss_s2}
\end{align}
where
\begin{align}
\vspace{-0.1cm}
  \mathcal{D}(x, \hat{x}) &= \gamma_{\mathrm{dist}} \|x - \hat{x}\|_2^2 + \gamma_s\, \mathcal{L}_{s}(x, \hat{x}), \\
  \mathcal{L}_{\mathrm{real}} &= \gamma_p\, \mathcal{L}_{p}(x, \hat{x}) + \gamma_{\mathrm{adv}}\, \mathcal{L}_{\mathrm{adv}} + \gamma_{\mathrm{FD}}\, \mathcal{L}_{\mathrm{FD}}, \\
  \mathcal{L}_{\mathrm{adv}} &= \mathbb{E}[\log \mathbf{D}_{\lambda,\beta}(x)] + \mathbb{E}[\log(1 - \mathbf{D}_{\lambda,\beta}(\hat{x}))]. \label{eq:loss_adv}
\end{align}
Here, $\mathcal{R}$ is the bitrate of $\hat{y}$ and $\hat{z}$, estimated in training by replacing quantization with additive uniform noise~\cite{balle2018variational}, and $\lambda$ is the rate control of~\cref{sec:method_rate}. $\mathcal{L}_{s}$ is the CLIP semantic loss~\cite{radford2021learning}, and the perceptual loss $\mathcal{L}_{p}$ is LPIPS~\cite{zhang2018unreasonable} in stage~1 and overlap-chunked edge-aware DISTS~\cite{ding2020image, zhang2026ultra} from stage~2 on. The discriminator $\mathbf{D}_{\lambda,\beta}$, conditioned on $(\lambda, \beta)$, maximizes $\mathcal{L}_{\mathrm{adv}}$, and $\mathcal{L}_{\mathrm{FD}}$ is the FD-Loss~\cite{yang2026representation}. The realism control $\beta$ (\cref{sec:method_realism}) also weights $\mathcal{L}_{\mathrm{real}}$, and both $\lambda$ and $\beta$ are sampled from continuous ranges during training. Stage~3 keeps $\mathcal{L}_{\mathrm{S2}}$ but trains on $1024 \times 1024$ crops, following the high-resolution fine-tuning of AEIC~\cite{zhang2026ultra}.

\vspace{-0.4cm}
\paragraph{Rate--Realism Conditioned Discriminator}
Our discriminator $\mathbf{D}$ consists of trainable heads~\cite{kumari2022ensembling} on a frozen DINOv2~\cite{oquab2024dinov2} backbone. Reconstructions at different operating points should differ, since a lower $\beta$ trades realism for fidelity and a lower rate leaves more detail to synthesize. We therefore condition every head on $(\lambda, \beta)$ with the projection discriminator of~\citet{miyato2018cgans}, using embeddings of $\lambda$ and $\beta$ (details in \cref{sec:supp_discriminator}).

\vspace{-0.4cm}
\paragraph{FD-Loss}
We adopt the FD-Loss of~\citet{yang2026representation}, which directly minimizes the Fr\'echet distance (FD) between feature statistics of generated and real images. Whereas they use it as the sole objective for post-training image generators, we add it as a complementary loss term. FD compares the mean $\mu$ and covariance $\Sigma$ of the reconstructions' features with those of real images, $\mu_r$ and $\Sigma_r$. As a mini-batch is far too small to estimate a covariance, $\mu$ and $\Sigma$ are computed over a queue of the last $N$ features, of which only the current mini-batch's carry gradient. For each feature $f$,
\begin{equation}
  \frac{\partial\, \mathcal{L}_{\mathrm{FD}}}{\partial f} \approx \frac{2}{N}\bigl(f - M(f)\bigr),
  \label{eq:fd_grad}
\end{equation}
where $M$ is the optimal-transport map from the reconstructions' feature distribution to that of real images (derivation in \cref{sec:supp_fd_loss}). This gradient falls as $1/N$, yet $N$ cannot simply be reduced as $d$-dimensional features need  $N \gg d$ for a nearly unbiased estimate~\cite{chong2020effectively}. At our $N = 40{,}000$, the FD gradient is thus a small fraction of the other losses' gradients.  Following VQGAN~\cite{esser2021taming}, we therefore rescale FD to the gradient of DISTS (see~\cref{sec:supp_fd_loss}).

\endgroup

\section{Experiments}
\label{sec:experiments}
\vspace{-0.2cm}

We report two models: \textbf{\modelname{}} and \textbf{\modelname{}-FD}; the latter being identical to the former except for the use of the FD-Loss (\cref{sec:method_training_objective}) during training. We first describe the experimental setup in~\cref{sec:exp_setup}, then compare both models with variable-rate and fixed-rate codecs in~\cref{sec:exp_baselines}, and finally ablate the context model, the rate conditioning and the realism conditioning of the denoiser in~\cref{sec:exp_ablation}.

\subsection{Experiment Setup}
\label{sec:exp_setup}
\vspace{-0.1cm}

\paragraph{Dataset}
We train on DF2K~\cite{lim2017enhanced}, the CLIC2020 Professional training and validation sets~\cite{toderici2020clic} and the first $10{,}000$ images of LSDIR~\cite{li2023lsdir} for a total of $14{,}119$ images whose shorter side is at least $512$ pixels. Following prior work on generative compression~\cite{mentzer2020high, muckley2023improving, zhang2026ultra}, we evaluate on the validation split of \textbf{DIV2K}~\cite{agustsson2017ntire}, the test split of \textbf{CLIC2020}~\cite{toderici2020clic}, and the \textbf{Kodak} dataset~\cite{kodak1993kodak}.

\vspace{-0.5cm}
\paragraph{Training Details}
We train both models in the three stages of~\cref{sec:method_training_objective} with AdamW~\cite{loshchilov2019decoupled} on two NVIDIA A40 GPUs, sampling $\lambda$ log-uniformly from $[0.05, 32]$ throughout all training stages. We train on $512 \times 512$ crops in stages~1 and 2, and on $1024 \times 1024$ crops in stage~3. In stage~1, we train for $350$k steps, use a batch size of $8$ and a learning rate of $10^{-4}$. In stage~2, we sample $\beta$ from $[0, 5.12]$, and train for $45$k steps with a batch size of $32$ and an initial learning rate of $5 \times 10^{-5}$. In stage~3, we fine-tune for $10$k steps with a batch size of $8$ and an initial learning rate of $2 \times 10^{-5}$, using the $4{,}807$ training images that are large enough for a $1024 \times 1024$ crop. \modelname{}-FD computes the FD-Loss as a weighted sum of three feature spaces, Inception-v3~\cite{szegedy2016rethinking}, SigLIP~2~\cite{tschannen2025siglip} and MAE~\cite{he2022masked}. More training details are provided in the supplementary~\cref{sec:supp_training_details}.

\vspace{-0.5cm}
\paragraph{Evaluation Metrics}
We measure distortion with PSNR. For perceptual quality, we use two full-reference metrics, LPIPS~\cite{zhang2018unreasonable} and DISTS~\cite{ding2020image}, and three distribution metrics: FID~\cite{heusel2017gans} and the Fr\'echet distances in the feature spaces of DINOv2~\cite{oquab2024dinov2} (FD-DINOv2) and CLIP~\cite{radford2021learning} (FD-CLIP). Following \citet{mentzer2020high}, we compute the distribution metrics on $256 \times 256$ patches of the full-resolution images. We only report PSNR, LPIPS and DISTS on Kodak since it is too small for reliable distribution metrics. The bitrate is measured in bits per pixel (bpp) from the size of the actual bitstream. Both \modelname{} models are evaluated at eight $\lambda$ values between $0.25$ and $32$, and at $\beta = 5.12$ ($\beta = 1.28$ on Kodak). Coding efficiency is compared by BD-rate~\cite{bjontegaard2001calculation}. Supplementary~\cref{sec:supp_eval_metrics} gives more details.

\subsection{Comparison with Baselines}
\label{sec:exp_baselines}
\vspace{-0.1cm}

\begin{table}[t]
  \centering
  \caption{\textbf{Ablation of the context model.} We report BD-rate (\%, lower is better) in terms of PSNR and LPIPS on Kodak and DIV2K, relative to \modelname{} without either the DOA scan or the cross-tap state carry. All three models are evaluated after stage~1 of training, at $\beta = 0$.}
  \vspace{-0.3cm}
  \label{tab:ablation_context}
  \small
  \setlength{\tabcolsep}{4pt}
  \begin{tabular}{@{}cccccc@{}}
    \toprule
    & & \multicolumn{2}{c}{Kodak} & \multicolumn{2}{c}{DIV2K} \\
    \cmidrule(lr){3-4} \cmidrule(l){5-6}
    DOA scan & \smash{\begin{tabular}[b]{@{}c@{}}Cross-tap\\ State Carry\end{tabular}} & PSNR & LPIPS & PSNR & LPIPS \\
    \midrule
    $\times$ & $\checkmark$ & $-$4.4 & $-$1.1 & +0.1 & $-$3.0 \\
    $\checkmark$ & $\checkmark$ & $-$5.9 & $-$11.5 & $-$5.2 & $-$14.8 \\
    \bottomrule
  \end{tabular}
  \vspace{-0.3cm}
\end{table}

We compare \modelname{} and \modelname{}-FD with the variable-rate codecs CoD-Diffc~\cite{jia2026cod}, Control-GIC~\cite{li2025once}, Diffc~\cite{vonderfecht2025lossy}, GLC~\cite{jia2024generative}, OSCAR~\cite{guo2025oscar} and Turbo-DDCM~\cite{vaisman2026turbo}, and with the fixed-rate codecs AEIC-SE~\cite{zhang2026ultra}, DiffEIC~\cite{li2024toward}, DLF~\cite{xue2025dlf}, MS-ILLM~\cite{muckley2023improving}, OneDC~\cite{xue2025one}, PerCo~\cite{careil2024towards, korber2024perco}, ResULIC~\cite{ke2025ultra}, StableCodec~\cite{zhang2025stablecodec} and VTM~\cite{bross2021overview}.

\vspace{-0.4cm}
\paragraph{Quantitative Comparison}
\Cref{fig:rd} presents the rate--distortion--perception curves. A single \modelname{} model spans the whole range from about $0.003$ to $0.06$\,bpp and clearly outperforms all \emph{variable-rate} codecs in every metric. Among all codecs, \modelname{} achieves the lowest DISTS on all three datasets at almost every bitrate and, together with \modelname{}-FD, the lowest FID on DIV2K and CLIC2020, and its LPIPS is second only to that of OneDC. The closest competitor across all metrics is AEIC-SE, whose five models stop below $0.04$\,bpp. Relative to AEIC-SE, \modelname{} saves $6.1\%$, $10.6\%$ and $7.2\%$ of the bitrate at equal LPIPS on DIV2K, Kodak and CLIC2020; $24.5\%$, $13.4\%$ and $22.3\%$ at equal DISTS; and $23.5\%$ and $19.9\%$ at equal FID on DIV2K and CLIC2020. \modelname{}-FD is better than \modelname{} in FD-DINOv2 and FD-CLIP and slightly worse in PSNR, LPIPS and DISTS, while the two are close in FID. More curves and BD-rates for all codecs are in \cref{sec:supp_quantitative_comparison}.

\vspace{-0.4cm}
\paragraph{Qualitative Comparison}
\Cref{fig:qual} shows two images coded by \modelname{} at $0.0106$ and $0.0028$\,bpp. Although it uses fewer bits than every baseline, \modelname{} achieves the best LPIPS and DISTS on both. On the DIV2K image, it keeps the crop rows and the tire tracks in the soil, which VTM and MS-ILLM blur and StableCodec smooths away. On the CLIC2020 image, it keeps the original color bands, which OneDC and OSCAR wash out.

\vspace{-0.4cm}
\paragraph{Computational Complexity}
Being variable-rate, \modelname{} needs far fewer parameters than the strongest fixed-rate codecs: a single $1.01$B-parameter model covers $256$ rate points, whereas AEIC-SE, OneDC and StableCodec need $4.4$--$11.3$B parameters for their $5$--$9$ rates. On a DIV2K subset, \modelname{} decodes $10\%$ faster than AEIC-SE but encodes $0.25$\,s slower, and decodes $1.5$--$1.9\times$ and encodes about $6\times$ faster than StableCodec and OneDC. Supplementary~\Cref{sec:supp_complexity} compares the complexity of all codecs.

\begin{table}[t]
  \centering
  \caption{\textbf{Ablation of the rate conditioning.}  We report BD-rate (\%, lower is better) in terms of PSNR and LPIPS on Kodak and DIV2K, relative to the full \modelname{}. All models are evaluated after stage~1 of training, at $\beta = 0$.}
  \vspace{-0.3cm}
  \label{tab:ablation_rate}
  \small
  \setlength{\tabcolsep}{3pt}
  \begin{tabular}{@{}lcccc@{}}
    \toprule
    & \multicolumn{2}{c}{Kodak} & \multicolumn{2}{c}{DIV2K} \\
    \cmidrule(lr){2-3} \cmidrule(l){4-5}
    Model & PSNR & LPIPS & PSNR & LPIPS \\
    \midrule
    \modelname{} w/o rate-adaptive scaling & +3.0 & +13.1 & +4.5 & +16.8 \\
    \modelname{} w/o CondStar blocks & +0.6 & +0.7 & +7.0 & +9.3 \\
    \bottomrule
  \end{tabular}
  \vspace{-0.2cm}
\end{table}
\begin{table}[t]
  \centering
  \caption{\textbf{Ablation of the realism conditioning of the denoiser.} We report BD-rate (\%, lower is better) of \modelname{} without the $\beta$-conditioned UNet on DIV2K, relative to the full \modelname{}.}
  \vspace{-0.3cm}
  \label{tab:ablation_realism}
  \small
  \setlength{\tabcolsep}{4pt}
  \begin{tabular}{@{}ccccccc@{}}
    \toprule
    $\beta$ & PSNR & LPIPS & DISTS & FID & FD-DINOv2 & FD-CLIP \\
    \midrule
    0.16 & +2.0 & $-$0.1 & +0.6 & $-$0.2 & $-$3.8 & $-$2.6 \\
    0.64 & +1.6 & +1.0 & +2.0 & +4.5 & +1.6 & +5.9 \\
    5.12 & +0.7 & +1.4 & +4.2 & +5.2 & +4.8 & +8.9 \\
    \bottomrule
  \end{tabular}
  \vspace{-0.4cm}
\end{table}

\subsection{Ablation Study}
\label{sec:exp_ablation}
\vspace{-0.1cm}

\paragraph{Context Model}
\Cref{tab:ablation_context} ablates the two components of the context model. Cross-tap state carry alone improves both PSNR and LPIPS on Kodak, but only LPIPS on DIV2K, where PSNR is unchanged. Adding the DOA scan raises the savings substantially, to $5$--$6\%$ at equal PSNR and $11$--$15\%$ at equal LPIPS.

\vspace{-0.4cm}
\paragraph{Rate Conditioning}
\Cref{tab:ablation_rate} removes each of the two rate-conditioning mechanisms in turn, and neither suffices on its own. Without rate-adaptive scaling, the bitrate rises by $3$--$5\%$ at equal PSNR and by $13$--$17\%$ at equal LPIPS on both datasets. Without the CondStar blocks, it rises by $7$--$9\%$ on DIV2K, but by less than $1\%$ on Kodak.

\vspace{-0.4cm}
\paragraph{Realism Conditioning on the Denoiser}
\Cref{tab:ablation_realism} analyzes a model without $\beta$ in the UNet, so that $\beta$ reaches the decoder only through $g_s$. Its PSNR penalty is largest at low $\beta$ ($2.0\%$ at $\beta = 0.16$), whereas its perceptual penalty grows with $\beta$, reaching $4$--$9\%$ in DISTS and FD metrics at $\beta = 5.12$. The ablated model is thus less faithful at low $\beta$ and less realistic at high $\beta$, which shows that conditioning the UNet on $\beta$ widens the distortion--perception trade-off.
\section{Conclusion}
\label{sec:conclusion}
\vspace{-0.1cm}

We introduced \modelname{}, a one-step diffusion codec in which a single model offers continuous control over rate, distortion and realism through two complementary knobs: $\lambda$ and $\beta$. Its VSS context model aligns the Mamba scan with the decoding order and carries its state across taps, so that every latent is predicted from a state that summarizes all decoded context. \modelname{} outperforms existing variable-rate codecs and even the strongest fixed-rate ones, with fewer total parameters. Like other generative codecs, it trades some fidelity for realism: its PSNR stays below that of distortion-oriented codecs, and at the lowest rates, it can synthesize unwanted details.

{
    \small
    \bibliographystyle{ieeenat_fullname}
    \bibliography{main}
}

\clearpage
\setcounter{page}{1}

\twocolumn[{%
  \centering
  \Large
  \textbf{\thetitle}\\
  \vspace{0.5em}Supplementary Material \\
  \vspace{1.0em}
  \normalsize
  \includegraphics[width=0.9\linewidth]{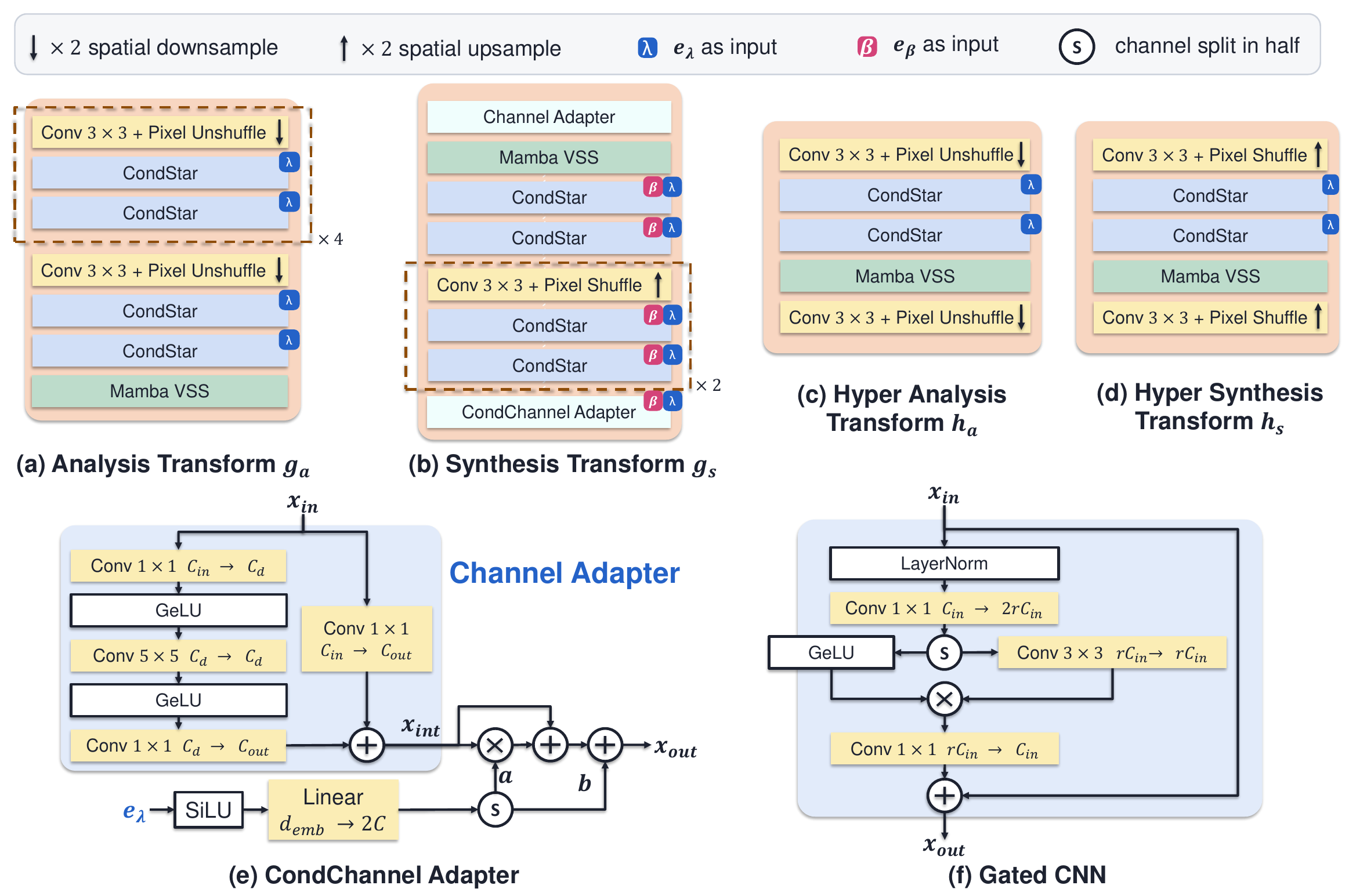}
  \captionof{figure}{\textbf{Transforms and other network components of \modelname{}.}
    (a)~Analysis transform $g_a$, whose dashed stage is repeated four times.
    (b)~Synthesis transform $g_s$, whose dashed stage is repeated twice.
    (c)~Hyper analysis transform $h_a$. 
    (d)~Hyper synthesis transform $h_s$.
    (e)~The CondChannel Adapter modulates the output $x_{\mathrm{int}}$ of the Channel Adapter (shaded) with a scale $\mathbf{a}$ and a shift $\mathbf{b}$ predicted from the embedding $\mathbf{e}_\lambda$. The Channel Adapter at the input of $g_s$ is the shaded part alone, \ie, $x_{\mathrm{out}} = x_{\mathrm{int}}$.
    (f)~Gated CNN block inside the context blocks of $g_c$ (\cref{fig:context_model}b), with expansion ratio $r$.
    The badges on a block mark the embeddings it receives, and a block with both badges receives $\mathbf{e}_\lambda + \mathbf{e}_\beta$. CondStar blocks are shown in \cref{fig:film_ras}b, and Mamba VSS blocks in \cref{fig:context_model}c.}
  \label{fig:supp_architecture}
  \vspace{1.0em}
}]

\appendix

\section{Network Architecture Details}
\label{sec:supp_network_details}

This section lists the configuration of the network blocks in \cref{fig:architecture}. \Cref{fig:supp_architecture} shows the transforms $g_a$, $g_s$, $h_a$ and $h_s$, the CondChannel Adapter and the Gated CNN block, and \cref{fig:context_model} the context model $g_c$.

\paragraph{Embeddings}
The $\lambda$- and $\beta$-embedding networks (\cref{fig:architecture}) scale $\lambda$ to $64 \log\lambda / \log\lambda_{\max}$ with $\lambda_{\max} = 32$, and $\beta$ to $64\,(\beta/\beta_{\max} - 0.5)$ with $\beta_{\max} = 5.12$. Each scaled value is mapped by a $256$-dimensional sinusoidal embedding with a maximum period of $64$, followed by an MLP (linear, SiLU, linear), to $\mathbf{e}_\lambda$ or $\mathbf{e}_\beta$ with $256$ dimensions. For entropy coding, $\lambda$ is snapped to one of $256$ values spaced logarithmically over $[0.05, 32]$ and sent as a one-byte index.

\paragraph{Transforms}
The analysis transform $g_a$ (\cref{fig:supp_architecture}a) has five stages with $64$, $128$, $192$, $256$ and $320$ channels, each a downsampling block followed by two CondStar blocks, and a VSS block after the last stage, so that $y$ has $M = 320$ channels at $1/32$ of the image resolution. The hyper analysis transform $h_a$ (\cref{fig:supp_architecture}c) consists of a downsampling block to $160$ channels, two CondStar blocks, a VSS block and a second downsampling block, so that $z$ has $160$ channels at $1/128$ of the image resolution. The hyper synthesis transform $h_s$ (\cref{fig:supp_architecture}d) has the same layout with upsampling blocks: the first keeps $160$ channels, and the second outputs $\varphi$ with $320$ channels at $1/32$ of the image resolution. The synthesis transform $g_s$ (\cref{fig:supp_architecture}b) begins with a Channel Adapter with $1024$ channels, a VSS block and two CondStar blocks, all at $1/32$ of the image resolution. Two stages with $768$ and $512$ channels follow, each made of an upsampling block and two CondStar blocks, and a final CondChannel Adapter outputs $576$ channels at $1/8$ of the image resolution, which are split into $l_T$ ($320$ channels) and $l_{\mathrm{res}}$ ($256$ channels). The conditioned blocks of $g_a$, $h_a$ and $h_s$ receive $\mathbf{e}_\lambda$, and those of $g_s$ receive $\mathbf{e}_\lambda + \mathbf{e}_\beta$.

\paragraph{Entropy Model}
$\hat{z}$ is coded with a factorized density~\cite{balle2018variational}, and $\hat{y}$ with a Gaussian whose scale is quantized to a table of $64$ values spaced logarithmically over $[0.11, 256]$. The context model $g_c$ (\cref{fig:context_model}a) has $N_c = 4$ context blocks with $3M = 960$ channels. In each context block (\cref{fig:context_model}b), the InceptionNeXt block~\cite{yu2024inceptionnext} has an MLP ratio of $2$. The Gated CNN block is shown in~\cref{fig:supp_architecture}f. On $C$ channels, it applies a layer normalization and a $1 \times 1$ convolution to $2rC$ channels with $r = 2$. The output is split into two halves, where it multiplies the GELU of one half with a depthwise $3 \times 3$ convolution of the other, and projects the product back to $C$ channels with a $1 \times 1$ convolution. A residual connection from the input to the output of the block is also added. The VSS blocks (\cref{fig:context_model}c) have a state size of $n=16$, an expansion ratio of $2$, two scan directions and four channel groups. The prefix tokenizer~(\cref{fig:context_model}c) pools the previous tap's output to $4 \times 4$ and projects it with a $1 \times 1$ convolution to $16$ tokens for each direction. The adapters $A_{\mathrm{in},t}$ and $A_{\mathrm{out},t}$ are CondChannel Adapters (\cref{fig:supp_architecture}e) conditioned on $\mathbf{e}_\lambda$: $A_{\mathrm{in},t}$ maps the $320$ channels of the context input to $960$, and $A_{\mathrm{out},t}$ maps the $960$ channels of the output of $g_c$ to the $640$ channels of $\mu_t$ and $\sigma_t$.

\paragraph{Other Network Components}
Downsampling and upsampling blocks of $g_a$, $g_s$, $h_a$ and $h_s$ consist of a $3 \times 3$ convolution followed by a pixel unshuffle or pixel shuffle by a factor of $2$. A CondStar block (\cref{fig:film_ras}b) modulates the output of a Star block $S$ with FiLM~\cite{perez2018film}, as in \cref{eq:condstar}. The Star block is adopted from StarNet~\cite{ma2024rewrite}. 
The Channel Adapter maps $C_{\mathrm{in}}$ to $C_{\mathrm{out}}$ channels using two parallel branches (shaded in \cref{fig:supp_architecture}e). The first branch is a single $1 \times 1$ convolution. The second branch applies a $1 \times 1$ convolution to $C_d = (C_{\mathrm{in}} + C_{\mathrm{out}})/2$ channels, a depthwise $5 \times 5$ convolution and a $1 \times 1$ convolution to $C_{\mathrm{out}}$ channels, with a GELU after each of the first two convolutions. The outputs of the two branches are summed into $x_{\mathrm{int}}$. The CondChannel Adapter modulates $x_{\mathrm{int}}$ in the same way as a CondStar block, $x_{\mathrm{out}} = (1 + \mathbf{a}) \odot x_{\mathrm{int}} + \mathbf{b}$ with $[\mathbf{a}, \mathbf{b}] = W\, \mathrm{SiLU}(\mathbf{e})$. In both conditioned blocks, $W$ is initialized to zero, so that every block starts unmodulated. The Mamba VSS blocks of the transforms are VSS blocks~\cite{liu2024vmamba} as in \cref{fig:context_model}c, but without the DOA token alignment and the prefix tokenizer. Each is applied after a layer normalization with a residual connection, and scans in raster order in two directions, with a state size of $16$ and an expansion ratio of $2$.

\paragraph{Denoiser and VAE Decoder}
The denoiser $\epsilon_{\mathrm{SD}}$ in~\cref{fig:architecture} is the UNet of SD-Turbo~\cite{sauer2024adversarial} without text conditioning, so its cross-attention layers are skipped. Its first convolution, which maps the $4$ channels of an SD latent to $320$, is widened to take the $320$ channels of $l_T$ by tiling its kernel $80$ times along the input channels and dividing it by $80$, and its last convolution is widened to output $256$ channels by tiling its $4$ output filters $64$ times. Both are initialized this way and then trained as full-size layers, while the remaining layers are frozen and fine-tuned with LoRA~\cite{hu2022lora} of rank $32$. $\beta$ enters the denoiser through the conditioning MLP $\mathcal{M}_{\mathrm{cond}}$ (\cref{fig:film_ras}c), detailed in \cref{sec:supp_cond_unet}. The VAE decoder $\mathcal{D}_{\mathrm{SD}}$ is the pruned, half-width decoder of \citet{chen2025adversarial}, which AEIC~\cite{zhang2026ultra} also uses, with its first convolution widened to $256$ input channels in the same way as the denoiser~$\epsilon_{\mathrm{SD}}$, and it stays frozen. Of the $1.01$\,B parameters of \modelname{}, $221$\,M are trained.

\section{Realism Control}
\label{sec:supp_cond_unet}

\begin{figure*}[t]
  \centering
  \begin{subfigure}{0.9\linewidth}
    \centering
    \includegraphics[width=\linewidth]{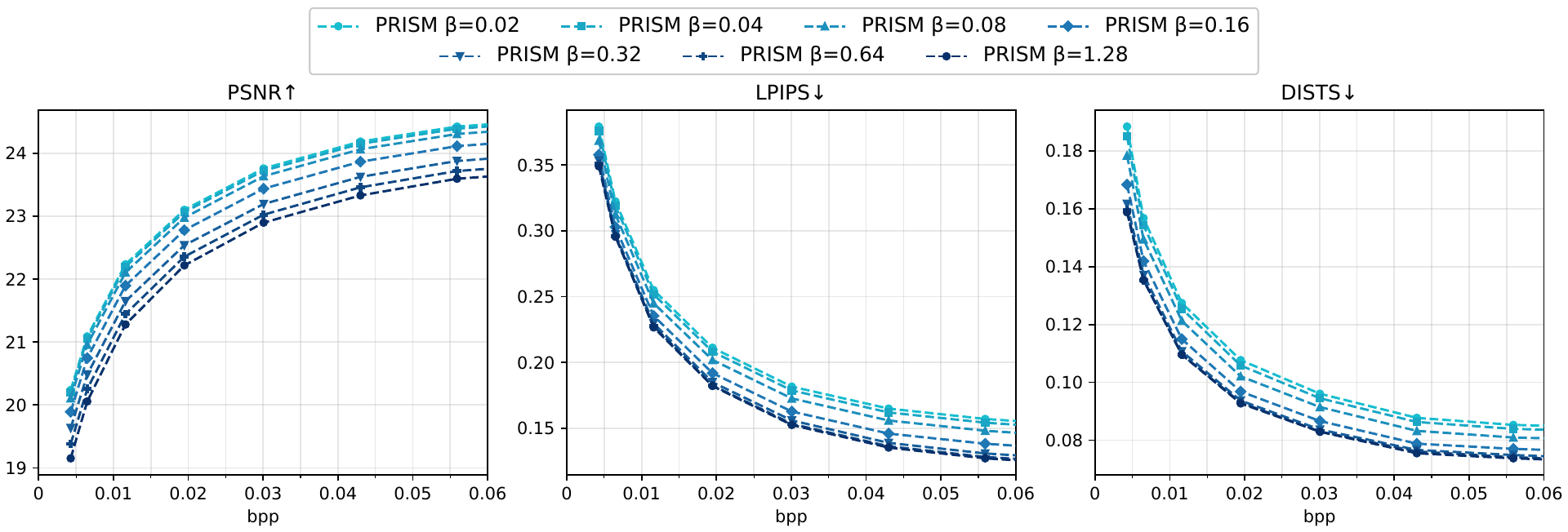}
    \caption{Kodak}
    \label{fig:rd_beta_kodak}
  \end{subfigure}
  \\[4pt]
  \begin{subfigure}{0.9\linewidth}
    \centering
    \includegraphics[width=\linewidth]{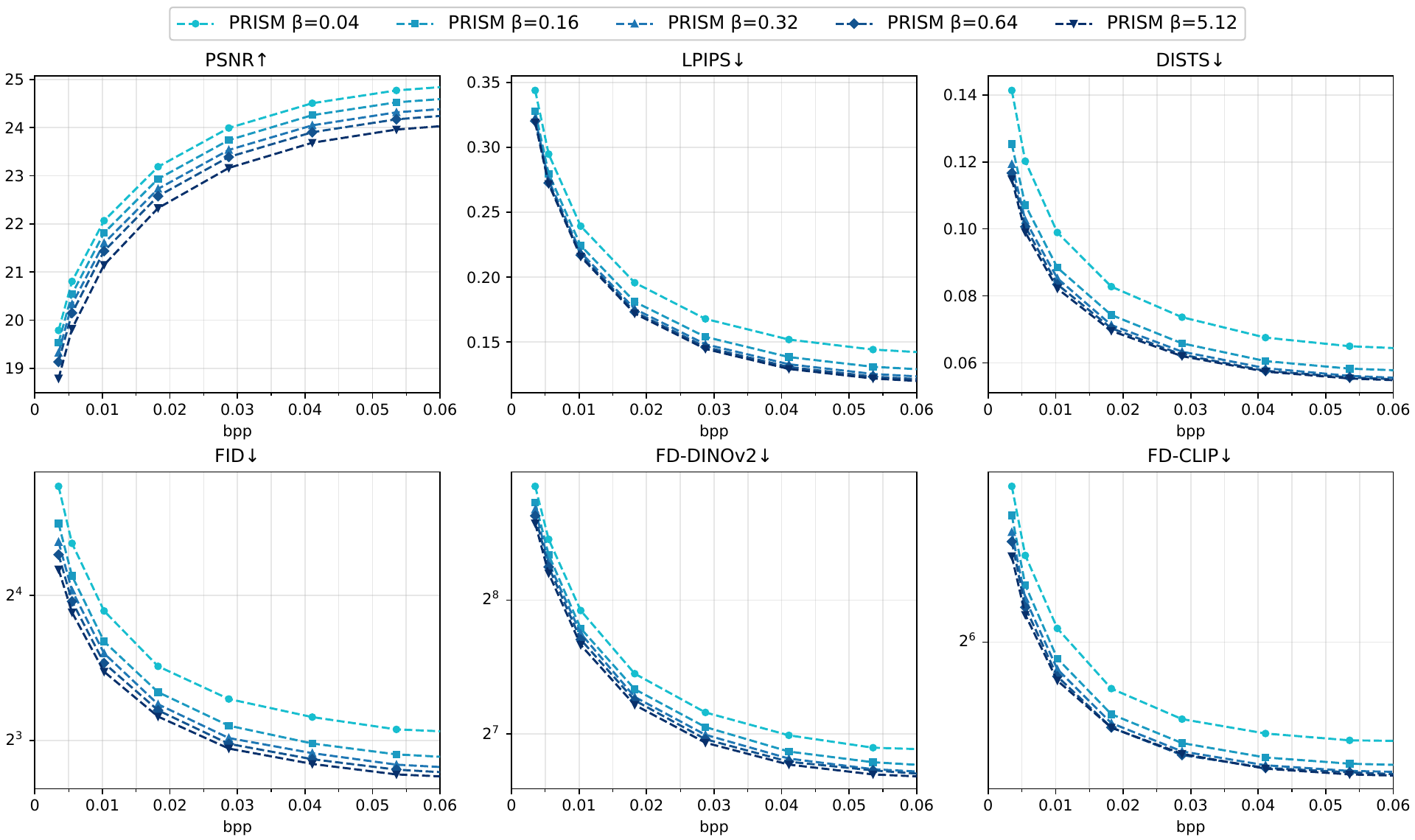}
    \caption{DIV2K}
    \label{fig:rd_beta_div2k}
  \end{subfigure}
  \caption{\textbf{Distortion--perception trade-off controlled by $\beta$} on Kodak and DIV2K. Each curve is \modelname{} decoded at one $\beta$ over the same eight values of $\lambda$. Since $\beta$ conditions only the decoder, the bitrate at a given $\lambda$ is the same for every $\beta$. FID, FD-DINOv2 and FD-CLIP are reported only on DIV2K, as Kodak is too small for distribution metrics.}
  \label{fig:rd_beta}
\end{figure*}

The realism control $\beta$ sets how much detail the decoder synthesizes (\cref{sec:method_realism}). This section first shows the distortion--perception trade-off that $\beta$ controls, and then details how $\mathcal{M}_{\mathrm{cond}}$ conditions the denoiser $\epsilon_{\mathrm{SD}}$ on $\beta$.

\paragraph{Distortion--Perception Trade-off}
Since a single \modelname{} model is trained over $\beta \in [0, 5.12]$, the trade-off is chosen at decoding time by $\beta$ alone. To measure it, we decode Kodak at seven values of $\beta$ from $0.02$ to $1.28$, and DIV2K at five values from $0.04$ to $5.12$, each at the eight values of $\lambda$ of \cref{sec:exp_setup}. \Cref{fig:rd_beta} shows the resulting curves. On Kodak, going from $\beta = 0.02$ to $1.28$ costs $72\%$ more bits at equal PSNR but saves $30\%$ and $38\%$ at equal LPIPS and DISTS. On DIV2K, going from $\beta = 0.04$ to $5.12$ costs $61\%$ more bits at equal PSNR but saves $29\%$ and $47\%$ at equal LPIPS and DISTS, and $49\%$, $29\%$ and $40\%$ at equal FID, FD-DINOv2 and FD-CLIP. On both datasets, the curves for intermediate values of $\beta$ lie in order between those of the lowest and the highest $\beta$, showing that $\beta$ moves the decoder gradually along the trade-off. 

\paragraph{Folding the UNet Denoiser Timestep}
Each of the $22$ residual blocks of the UNet of SD-Turbo adds a projection of the timestep embedding $\mathbf{e}_T$ to the output of its first convolution,
\begin{equation}
  f \leftarrow \mathrm{conv}_1(f) + P\bigl(\mathrm{SiLU}(\mathbf{e}_T)\bigr),
  \label{eq:supp_temb}
\end{equation}
where $P$ is an affine projection of the block. Since \modelname{} always uses $T = 999$, the added term is a constant, and we fold it into the bias of $\mathrm{conv}_1$. 

\paragraph{Realism Conditioned Timestep Offset}
$\mathcal{M}_{\mathrm{cond}}$ (\cref{fig:film_ras}c) maps $\mathbf{e}_\beta$ to an offset $\delta \in \mathbb{R}^{1280}$ of $\mathbf{e}_T$. Writing the projection in \cref{eq:supp_temb} as $P(x) = W_P\, x + b_P$, with weight matrix $W_P$ and bias $b_P$, each residual block of SD-Turbo computes
\begin{equation}
  h_{\mathrm{old}} = \mathrm{conv}_1(f) + W_P\, \mathrm{SiLU}(\mathbf{e}_T) + b_P,
  \label{eq:supp_old_h}
\end{equation}
where $W_P\, \mathrm{SiLU}(\mathbf{e}_T) + b_P$ is the term folded into the bias of $\mathrm{conv}_1$. At the offset timestep embedding $\mathbf{e}_T + \delta$, the block should instead compute
\begin{equation}
  h_{\mathrm{new}} = \mathrm{conv}_1(f) + W_P\, \mathrm{SiLU}(\mathbf{e}_T + \delta) + b_P.
  \label{eq:supp_new_h}
\end{equation}
The two differ only in the bias, by $h_{\mathrm{new}} - h_{\mathrm{old}} = W_P\, u$ with
\begin{equation}
  u = \mathrm{SiLU}(\mathbf{e}_T + \delta) - \mathrm{SiLU}(\mathbf{e}_T).
  \label{eq:supp_u}
\end{equation}
Before each pass of the UNet, we therefore compute $u$, and every residual block adds $W_P\, u$ to the output of its folded first convolution. Together with the folded bias, every block then computes exactly $h_{\mathrm{new}}$, \ie, what it would compute in SD-Turbo at the timestep embedding $\mathbf{e}_T + \delta$. The projections $P$ stay frozen and are adapted with LoRA of rank $8$; since the folded bias is fixed, the LoRA changes only the added term $W_P\, u$. The last layer of $\mathcal{M}_{\mathrm{cond}}$ is initialized to zero, so that at the start of stage~2, $\delta = 0$ and the decoder is exactly the one trained in stage~1. $\mathcal{M}_{\mathrm{cond}}$ and the LoRA of the projections add $2.36$\,M parameters.

\section{Context Model}
\label{sec:supp_context_model}

This section compares the context model $g_c$ of \modelname{}~(\cref{sec:method_context_model}) with those of existing codecs, ablates the number of its scan directions, and measures the contribution of each component of the context model. For the comparisons in \cref{tab:supp_ctx_models,tab:supp_ctx_directions}, we replace only the context model, train each model with stage~1 settings, and evaluate it at $\beta = 0$ and the eight values of $\lambda$ of~\cref{sec:exp_setup}. BD-rates are relative to \modelname{}.

\begin{table}[t]
  \centering
  \caption{\textbf{Comparison with existing context models.} BD-rate (\%, lower is better) in terms of PSNR and LPIPS, relative to \modelname{}, when only its context model is replaced. Params: parameters of the context model, including its adapters, in millions. All models are evaluated at $\beta = 0$.}
  \vspace{-0.2cm}
  \label{tab:supp_ctx_models}
  \small
  \setlength{\tabcolsep}{4pt}
  \begin{tabular}{@{}lccccc@{}}
    \toprule
    & & \multicolumn{2}{c}{Kodak} & \multicolumn{2}{c}{DIV2K} \\
    \cmidrule(lr){3-4} \cmidrule(l){5-6}
    Context model & Params & PSNR & LPIPS & PSNR & LPIPS \\
    \midrule
    AEIC~\cite{zhang2026ultra} & 59.9 & +2.7 & +2.9 & +8.2 & +8.3 \\
    MambaIC~\cite{zeng2025mambaic} & 99.5 & +3.0 & +3.7 & +8.6 & +5.9 \\
    \midrule
    \modelname{} & 97.8 & \multicolumn{4}{c}{reference} \\
    \bottomrule
  \end{tabular}
\end{table}

\begin{table}[t]
  \centering
  \caption{\textbf{Number of scan directions in the VSS blocks of $g_c$.} BD-rate (\%, lower is better) in terms of PSNR and LPIPS, relative to \modelname{}, which scans the DOA order in both directions. Params: parameters of the context model, including its adapters, in millions. Both models are evaluated at $\beta = 0$.}
  \vspace{-0.2cm}
  \label{tab:supp_ctx_directions}
  \small
  \setlength{\tabcolsep}{4pt}
  \begin{tabular}{@{}lccccc@{}}
    \toprule
    & & \multicolumn{2}{c}{Kodak} & \multicolumn{2}{c}{DIV2K} \\
    \cmidrule(lr){3-4} \cmidrule(l){5-6}
    Directions & Params & PSNR & LPIPS & PSNR & LPIPS \\
    \midrule
    Forward only & 90.0 & +2.9 & +9.3 & +3.0 & +5.9 \\
    \midrule
    Both (\modelname{}) & 97.8 & \multicolumn{4}{c}{reference} \\
    \bottomrule
  \end{tabular}
\end{table}

\paragraph{Comparison with Existing Context Models}
\Cref{tab:supp_ctx_models} replaces the context model of \modelname{} with those of AEIC~\cite{zhang2026ultra} and MambaIC~\cite{zeng2025mambaic}. For AEIC, we remove the VSS blocks from $g_c$, so that each of its four context blocks consists of the InceptionNeXt and Gated CNN blocks only, as in AEIC. The four taps, the $3M$ channels and the $\lambda$-conditioned adapters $A_{\mathrm{in},t}$ and $A_{\mathrm{out},t}$ stay the same. For MambaIC, we port its context model, which codes $\hat{y}$ in five channel slices, each in two checkerboard halves~\cite{he2021checkerboard}, \ie, in $10$ sequential steps instead of $4$ taps. Each step combines window attention, two VSS blocks that scan along the four routes of VMamba (\cref{sec:background_mamba}), and latent residual prediction~\cite{minnen2020channel}. We condition the VSS blocks on $\mathbf{e}_\lambda$ through a $\lambda$-conditioned adapter at their input and a FiLM layer on the means and scales of every step, in place of $A_{\mathrm{in},t}$ and $A_{\mathrm{out},t}$. Both replacements need more bits than \modelname{} in both metrics on both datasets: $2.7$--$3.0\%$ more at equal PSNR on Kodak, and $8.2$--$8.6\%$ more on the larger images of DIV2K. MambaIC falls behind although it has about as many parameters as \modelname{} and more than twice as many sequential steps.

\paragraph{Number of Scan Directions}
The VSS blocks of $g_c$ scan the DOA order in both directions (\cref{sec:method_context_model}). \Cref{tab:supp_ctx_directions} compares this with a single forward scan, in which the decoded tokens still come before the targets, and the backward scan is removed. A single direction saves half of the scans and $7.8$\,M parameters. It needs about $3.0\%$ more bits at equal PSNR, and $5.9$--$9.3\%$ more bits at equal LPIPS.

\begin{table}[t]
  \centering
  \caption{\textbf{Contribution of each component of the context model.} Increase (\%) in the bits of $\hat{y}$ when one component of the context model of \modelname{} is switched off at inference, with the latent $y$ and the hyperprior fixed, averaged over the eight values of $\lambda$. Total is the increase in the bits of all four taps together; $^\dagger$ marks the increase in the bits of taps~1--3 together, since switching off this component does not change tap~0.}
  \vspace{-0.2cm}
  \label{tab:supp_ctx_components}
  \small
  \setlength{\tabcolsep}{4pt}
  \begin{tabular}{@{}lccccc@{}}
    \toprule
    Switched off & Tap 0 & Tap 1 & Tap 2 & Tap 3 & Total \\
    \midrule
    \multicolumn{6}{@{}l}{\emph{Kodak}} \\
    DOA scan order & 0.0 & +1.5 & +1.3 & +1.5 & +1.5$^\dagger$ \\
    Position-specific context & +3.8 & +6.0 & +5.4 & +6.4 & +5.3 \\
    Previous-tap summary & 0.0 & +0.5 & +1.2 & +1.9 & +1.1$^\dagger$ \\
    \midrule
    \multicolumn{6}{@{}l}{\emph{DIV2K}} \\
    DOA scan order & 0.0 & +2.0 & +7.2 & +1.4 & +3.4$^\dagger$ \\
    Position-specific context & +3.2 & +7.9 & +8.7 & +9.3 & +6.5 \\
    Previous-tap summary & 0.0 & +0.2 & +0.4 & +0.6 & +0.4$^\dagger$ \\
    \bottomrule
  \end{tabular}
\end{table}

\paragraph{Contribution of Each Component}
To see which components the trained context model relies on, we switch them off at inference~(\cref{tab:supp_ctx_components}). We take \modelname{} after the full $350$k steps of stage~1, encode each image once per $\lambda$, and rerun only the four taps of $g_c$ with one component switched off, so that the latent $y$ and the hyperprior are those of the intact model. Each variant quantizes $y$ around its own predicted means and passes its own $\hat{y}$ on to the later taps. We switch off three components:
(i)~the DOA scan order, by scanning in raster order and its reverse instead;
(ii)~the position-specific context, by replacing the readout $C_j h_j$ of the state in \cref{eq:ssm_output} with its mean over all positions $j$, per image and channel, in both directions, which keeps the average context but removes how it varies between positions;
(iii)~the previous-tap summary, by setting the prefix tokens of both directions to zero, so that every scan starts from an empty state, as at tap~0.
At tap~0, the scan is already in raster order and there is no previous tap, so (i) and (iii) change only taps~1--3.

Removing the position-specific context costs the most, $5.3\%$ of the bits on Kodak and $6.5\%$ on DIV2K. Its cost is $5.4$--$9.3\%$ at taps~1--3, where decoded latents provide context, compared with $3.2$--$3.8\%$ at tap~0. Scanning in raster order instead of the DOA order costs $1.5\%$ on Kodak and $3.4\%$ on DIV2K at taps~1--3, so the trained model depends on reading the decoded latents in the DOA order. The previous-tap summary saves less, $1.1\%$ on Kodak and $0.4\%$ on DIV2K at taps~1--3, and its saving rises from tap~1 to tap~3 on both datasets. Note that these results show how much the trained model relies on each component for entropy coding, not how much each component contributes to the training of the model, which the models retrained without them in~\cref{tab:ablation_context} measure.

\section{Rate--Realism Conditioned Discriminator}
\label{sec:supp_discriminator}

The discriminator $\mathbf{D}_{\lambda,\beta}$ follows the vision-aided discriminator of \citet{kumari2022ensembling} with a DINOv2 backbone, and conditions its heads on $(\lambda, \beta)$.

\paragraph{Backbone and Heads}
Each real or reconstructed training crop, $512 \times 512$ in stage~2 and $1024 \times 1024$ in stage~3, is resized as a whole to $224 \times 224$ and passed through a frozen DINOv2 ViT-B/14~\cite{oquab2024dinov2}. Three trainable heads read its features. Two convolutional heads take the patch tokens of the 5th and 9th of its $12$ blocks, and each predicts a $4 \times 4$ map of logits, one for each region of the image. An MLP head takes the class token of the last block and predicts a single logit for the whole image. Each head maps its input to $128$ hidden channels $\mathbf{v}_i$ before its output layer $\psi_i$. As in \citet{kumari2022ensembling}, real and reconstructed images are augmented with random color changes, translations and cutouts before the backbone.

\paragraph{Conditioning on $\lambda$ and $\beta$}
The discriminator has its own $\lambda$- and $\beta$-embedding networks, with the same architecture and scaling as those of the codec (\cref{sec:supp_network_details}), and sums their outputs into a condition $\mathbf{c}$. Following the projection discriminator~\cite{miyato2018cgans}, each head $i$ adds the inner product of its hidden features with a linear projection $\Pi_i$ of $\mathbf{c}$ to its logit,
\begin{equation}
  \mathbf{D}_{\lambda,\beta}^{(i)}(x) = \psi_i(\mathbf{v}_i) + \tfrac{1}{\sqrt{128}}\, \langle \Pi_i\, \mathbf{c},\, \mathbf{v}_i \rangle,
  \label{eq:supp_disc_proj}
\end{equation}
where, for the two convolutional heads, the inner product is taken at every position of the logit map. The condition thus changes which features count as real at each operating point. The projections $\Pi_i$ are initialized to zero, so that the discriminator starts unconditioned. The discriminator uses its own embeddings rather than $\mathbf{e}_\lambda$ and $\mathbf{e}_\beta$ of the codec. Otherwise, the gradient of the adversarial loss would also flow into the embedding networks of the codec, which could then fool the discriminator by changing the conditioning instead of the reconstruction. The heads, embeddings and projections have $2.23$\,M trainable parameters in total. The loss with which the discriminator is trained is given in \cref{sec:supp_training_details}.

\section{Adaptive Fr\'echet Distance Loss (FD-Loss)}
\label{sec:supp_fd_loss}

\begin{figure}[t]
  \centering
  \includegraphics[width=0.8\linewidth]{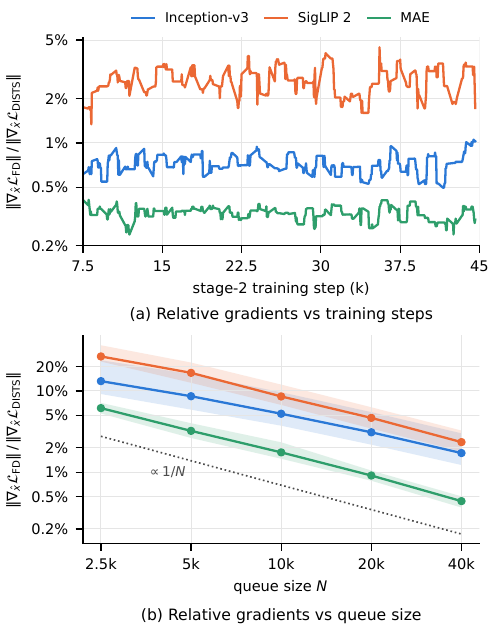}
  \caption{\textbf{Gradient of FD before rescaling, relative to that of DISTS}, both measured on the reconstruction $\hat{x}$, for the three backbones of \modelname{}-FD; this is $1/s_k$ in \cref{eq:fd_adaptive}.
    (a)~Over stage~2 of training, from step $7{,}500$, where FD is switched on, as a moving median over $9$ logged steps.
    (b)~At the final checkpoint, against the size $N$ of a queue filled with features of the model's own reconstructions at $\beta = 5.12$: median over $16$ held-out $1024 \times 1024$ crops, with the interquartile range shaded. The dotted line has slope $-1$.}
  \label{fig:fd_gradient}
\end{figure}

This section derives the gradient of the FD-Loss in \cref{eq:fd_grad}, explains why it is small when FD is one term among several losses, and defines the adaptive weight with which we rescale it.

\paragraph{FD Estimated from a Queue}
Let $\Phi$ denote a frozen backbone, which maps a $256 \times 256$ image patch to a $d$-dimensional feature. FD models the features of real patches $x$ and of reconstructed patches $\hat{x}$ as multivariate Gaussians, $\mathcal{N}(\mu_r, \Sigma_r)$ and $\mathcal{N}(\mu, \Sigma)$, with means and covariances
\begin{equation}
  \begin{aligned}
    \mu_r &= \mathbb{E}[\Phi(x)], & \Sigma_r &= \operatorname{Cov}[\Phi(x)], \\
    \mu &= \mathbb{E}[\Phi(\hat{x})], & \Sigma &= \operatorname{Cov}[\Phi(\hat{x})].
  \end{aligned}
  \label{eq:supp_fd_stats}
\end{equation}
The FD between the two Gaussians is usually written as~\cite{dowson1982frechet, yang2026representation}
\begin{equation}
  \mathrm{FD} = \|\mu - \mu_r\|^2 + \operatorname{Tr}\bigl(\Sigma_r + \Sigma - 2 (\Sigma_r \Sigma)^{1/2}\bigr),
  \label{eq:supp_fd_common}
\end{equation}
where $\operatorname{Tr}(\cdot)$ is the trace of a square matrix, the sum of its diagonal entries, which also equals the sum of its eigenvalues. For a symmetric positive semi-definite matrix $S$, $S^{1/2}$ denotes the symmetric positive semi-definite matrix with $S^{1/2} S^{1/2} = S$. The product $\Sigma_r \Sigma$ is in general not symmetric, and $(\Sigma_r \Sigma)^{1/2}$ denotes its principal square root, the one whose eigenvalues are non-negative.

We now rewrite \cref{eq:supp_fd_common} with a symmetric matrix inside the square root. Since the trace is linear, its trace term splits into $\operatorname{Tr}(\Sigma) + \operatorname{Tr}(\Sigma_r) - 2 \operatorname{Tr}\bigl((\Sigma_r \Sigma)^{1/2}\bigr)$, so only the last term needs rewriting. Let $\Sigma_r$ be positive definite, so that $\Sigma_r^{-1/2}$ exists, and let $A = \Sigma_r^{1/2} \Sigma\, \Sigma_r^{1/2}$, which is symmetric. Writing $\Sigma_r$ as $\Sigma_r^{1/2} \Sigma_r^{1/2}$, and inserting $\Sigma_r^{1/2} \Sigma_r^{-1/2} = I$ after $\Sigma$ gives
\begin{equation}
    \begin{aligned}
        \Sigma_r \Sigma &= \Sigma_r^{1/2} \bigl(\Sigma_r^{1/2} \Sigma\, \Sigma_r^{1/2}\bigr) \Sigma_r^{-1/2} \\
        &= \Sigma_r^{1/2} A\, \Sigma_r^{-1/2}, \\
        \Rightarrow (\Sigma_r \Sigma)^{1/2} &= \Sigma_r^{1/2} A^{1/2}\, \Sigma_r^{-1/2}, \\
        \Rightarrow \operatorname{Tr}\bigl((\Sigma_r \Sigma)^{1/2}\bigr) &= \operatorname{Tr}\bigl(\Sigma_r^{1/2} A^{1/2}\, \Sigma_r^{-1/2}\bigr).
    \end{aligned}
\end{equation}
Since $\operatorname{Tr}(XY) = \operatorname{Tr}(YX)$ for any two matrices $X$ and $Y$,
\begin{equation}
  \begin{aligned}
    \operatorname{Tr}\bigl((\Sigma_r \Sigma)^{1/2}\bigr) &= \operatorname{Tr}\bigl(A^{1/2}\, \Sigma_r^{-1/2} \Sigma_r^{1/2}\bigr) \\
    &= \operatorname{Tr}(A^{1/2}).
  \end{aligned}
\end{equation}
Substituting this into \cref{eq:supp_fd_common} gives
\begin{equation}
    \begin{aligned}
          \mathrm{FD} = \|\mu - \mu_r\|^2 &+ \operatorname{Tr}(\Sigma) + \operatorname{Tr}(\Sigma_r) \\
          &- 2 \operatorname{Tr}\bigl((\Sigma_r^{1/2}\, \Sigma\, \Sigma_r^{1/2})^{1/2}\bigr).
          \label{eq:supp_fd}
    \end{aligned}
\end{equation}
We use the form of~\cref{eq:supp_fd} with the term $\operatorname{Tr}(A^{1/2})$, which is the sum of the square roots of the eigenvalues of the symmetric matrix $A$, to derive the gradient of the FD-Loss.

The reference statistics $\mu_r$ and $\Sigma_r$ are computed once, on $50{,}000$ random $256 \times 256$ crops of our training images. The statistics of the reconstructions are estimated from a queue of the last $N$ features of reconstructed patches, $f_i = \Phi(\hat{x}_i)$ for $i = 1, \dots, N$,
\begin{equation}
  \mu = \frac{1}{N} \sum_{i=1}^{N} f_i, \qquad \Sigma = \frac{1}{N - 1} \sum_{i=1}^{N} (f_i - \mu)(f_i - \mu)^\top.
\end{equation}
At every step, the features of the current mini-batch replace the oldest entries of the queue. Only these entries depend on the current parameters, while the others were computed at earlier steps and carry no gradient.

\paragraph{Gradient with Respect to One Feature}
Let $f_j$ be a feature of the current mini-batch. The mean term gives
\begin{equation}
  \frac{\partial \|\mu - \mu_r\|^2}{\partial f_j} = \Bigl(\frac{\partial \mu}{\partial f_j}\Bigr)^{\!\top} 2 (\mu - \mu_r) = \frac{2}{N} (\mu - \mu_r),
\end{equation}
since $\partial \mu / \partial f_j = I / N$. 
For the covariance terms, we use $A = \Sigma_r^{1/2} \Sigma\, \Sigma_r^{1/2}$ from~\cref{eq:supp_fd} and let $T = \Sigma_r^{1/2} A^{-1/2}\, \Sigma_r^{1/2}$. For a positive-definite $A$,
\begin{equation}
  \frac{\partial \operatorname{Tr}(A^{1/2})}{\partial A} = \tfrac{1}{2} A^{-1/2}.
  \label{eq:supp_dtrsqrt}
\end{equation}
Since every entry of $A$ is a linear combination of entries of $\Sigma$ and $\Sigma_r^{1/2}$ is symmetric, the chain rule through $A$ places $\Sigma_r^{1/2}$ on each side of the derivative with respect to $A$. With \cref{eq:supp_dtrsqrt} and $\partial \operatorname{Tr}(\Sigma) / \partial \Sigma = I$,
\begin{equation}
  \begin{aligned}
    & \frac{\partial \operatorname{Tr}(A^{1/2})}{\partial \Sigma} = \Sigma_r^{1/2}\, \frac{\partial \operatorname{Tr}(A^{1/2})}{\partial A}\, \Sigma_r^{1/2} \\
    \Rightarrow\; & \frac{\partial \operatorname{Tr}(A^{1/2})}{\partial \Sigma} = \tfrac{1}{2}\, \Sigma_r^{1/2} A^{-1/2}\, \Sigma_r^{1/2} = \tfrac{1}{2}\, T \\
    \Rightarrow\; & \frac{\partial}{\partial \Sigma} \Bigl[\operatorname{Tr}(\Sigma) - 2 \operatorname{Tr}(A^{1/2})\Bigr] = I - 2 \cdot \tfrac{1}{2}\, T = I - T.
  \end{aligned}
\end{equation}
$T$ is symmetric and positive definite, and $T \Sigma T = \Sigma_r^{1/2} A^{-1/2} A A^{-1/2} \Sigma_r^{1/2} = \Sigma_r$.

Since the covariance terms depend on $f_j$ only through $\Sigma$, the chain rule lets us hold $G = I - T$ fixed:
\begin{equation}
  \frac{\partial \bigl[\operatorname{Tr}(\Sigma) - 2 \operatorname{Tr}(A^{1/2})\bigr]}{\partial f_j} = \frac{\partial \operatorname{Tr}(G \Sigma)}{\partial f_j}.
\end{equation}
Using $\operatorname{Tr}(G\, v v^\top) = v^\top G\, v$, $\partial (v^\top G\, v) / \partial v = 2 G v$ for a symmetric $G$, and $\partial \mu / \partial f_j = I / N$,
\begin{equation}
  \begin{aligned}
    \frac{\partial \operatorname{Tr}(G \Sigma)}{\partial f_j}
    &= \frac{1}{N-1}\, \frac{\partial}{\partial f_j} \sum_{i=1}^{N} (f_i - \mu)^\top G\, (f_i - \mu) \\
    &= \frac{2}{N-1}\, G \Bigl[(f_j - \mu) - \frac{1}{N} \sum_{i=1}^{N} (f_i - \mu)\Bigr] \\
    &= \frac{2}{N-1}\, G\, (f_j - \mu),
  \end{aligned}
\end{equation}
since $\sum_i (f_i - \mu) = 0$. Adding the mean term and using $N - 1 \approx N$,
\begin{equation}
  \begin{aligned}
    \frac{\partial\, \mathrm{FD}}{\partial f_j}
    &= \frac{2}{N} (\mu - \mu_r) + \frac{2}{N - 1} (I - T)(f_j - \mu) \\
    &\approx \frac{2}{N} \bigl[f_j - \mu_r - T (f_j - \mu)\bigr] = \frac{2}{N} \bigl(f_j - M(f_j)\bigr),
  \end{aligned}
\end{equation}
with $M(f) = \mu_r + T (f - \mu)$, which is \cref{eq:fd_grad}. Since $T$ is symmetric positive definite and $T \Sigma T = \Sigma_r$, $M$ maps $\mathcal{N}(\mu, \Sigma)$ to $\mathcal{N}(\mu_r, \Sigma_r)$, and, as the gradient of a convex function, it is the optimal-transport map between these two Gaussians~\cite{dowson1982frechet}.

\paragraph{Magnitude of the FD-Loss Gradient}
When a loss is averaged over a mini-batch of $B$ images, each image receives a weight of $1/B$, so the weights of all images that carry gradient sum to $1$. FD behaves differently. By \cref{eq:fd_grad}, it gives each of the $N$ features in the queue a weight of $2/N$, but only the $b$ features of the current mini-batch carry gradient, so their weights sum to only $2b/N$. With $N = 40{,}000$ and $b = 8$ patches per GPU in stage~2 and $b = 16$ in stage~3, this total weight is only $4 \times 10^{-4}$ and $8 \times 10^{-4}$, respectively. This weight applies to the gradient with respect to the features. The gradient that finally reaches the reconstruction $\hat{x}$ is further scaled by how strongly the features of the backbone change with the pixels, and this differs from one backbone to another.

Simply reducing $N$ is not an option. A covariance estimated from $N$ samples has rank at most $N - 1$, so features of dimension $d$ require $N > d$, where $d$ is $2048$ for Inception-v3, $1152$ for SigLIP~2 and $1024$ for MAE. Even when $N > d$, the estimated FD is biased upward by a term that decays as $1/N$~\cite{chong2020effectively}. The large queue that makes FD a reliable estimate is therefore exactly what makes its gradient small.

Whether this small gradient matters depends on the other losses. When FD is the only loss, as in~\citet{yang2026representation}, AdamW cancels the factor $2/N$, because it divides the gradient by its running root mean square. In a sum of losses, however, AdamW normalizes the combined gradient, so each loss influences the update in proportion to the norm of its own gradient. FD then controls only a small part of the update.

Our measurements confirm this. Over stage~2 (\cref{fig:fd_gradient}a), the median gradient of FD on $\hat{x}$ is only $0.71\%$, $2.6\%$ and $0.34\%$ of that of DISTS for Inception-v3, SigLIP~2 and MAE, respectively, so the three backbones differ by a factor of about $8$. A fixed FD weight would therefore have to be tuned for each backbone. In~\cref{fig:fd_gradient}b, we test how the magnitude of the gradient of FD relative to DISTS changes with the queue size $N$. We can see the ratio decreases with $N$ with a slope between $-0.74$ and $-0.94$. This is close to the slope of $-1$ expected from the factor $2/N$.

\paragraph{Gradient Rescaling}
Every loss reaches the parameters $\theta$ of the codec through the reconstruction, $\nabla_\theta \mathcal{L} = (\partial \hat{x} / \partial \theta)^\top \nabla_{\hat{x}} \mathcal{L}$, where $\partial \hat{x} / \partial \theta$ is the same for every loss. Following VQGAN~\cite{esser2021taming}, we therefore balance the losses on $\hat{x}$ and rescale FD to the gradient of DISTS:
\begin{equation}
  \mathcal{L}_{\mathrm{FD}} = s_0\, \frac{1}{K} \sum_{k=1}^{K} s_k\, \mathrm{FD}_k, \quad
  s_k = \mathrm{sg}\!\left(\frac{\|\nabla_{\hat{x}} \mathcal{L}_{\mathrm{DISTS}}\|}{\|\nabla_{\hat{x}} \mathrm{FD}_k\|}\right),
  \label{eq:fd_adaptive}
\end{equation}
where $\mathrm{sg}$ is the stop-gradient and $\mathrm{FD}_k$ is the FD in the feature space of the $k$-th of $K = 3$ backbones. Each $s_k$, clamped at $10^4$, gives the gradient of $s_k \mathrm{FD}_k$ the norm of that of DISTS. Gradients that point in different directions partly cancel in the average, so a second factor $s_0$ restores the norm again, which makes $\gamma_{\mathrm{FD}}$ the actual ratio between the two losses. $s_0$ is defined as
\begin{equation}
    s_0 = \mathrm{sg}\left(\frac{\|\nabla_{\hat{x}} \mathcal{L}_{\mathrm{DISTS}}\|} { \|\frac{1}{K} \sum_k s_k \nabla_{\hat{x}} \mathrm{FD}_k\|}\right).
\end{equation}
The weights need no extra backward pass through the codec, since the gradients on $\hat{x}$ pass only through the frozen backbones and the DISTS network. As in VQGAN, which matches norms at the last layer of its decoder, the losses are balanced on $\hat{x}$ and only approximately on $\theta$, since $\partial \hat{x} / \partial \theta$ can amplify some directions of $\hat{x}$ more than others.

\section{Training Details}
\label{sec:supp_training_details}

This section lists the settings with which \modelname{} and \modelname{}-FD are trained (\cref{tab:supp_training}).

\begin{table}[t]
  \centering
  \caption{\textbf{Settings of the three training stages.} The batch size is the global batch, split over two GPUs and gradient accumulation; the learning rate is lowered in steps from its initial to its final value.}
  \label{tab:supp_training}
  \small
  \begin{tabular}{@{}lccc@{}}
    \toprule
    & Stage~1 & Stage~2 & Stage~3 \\
    \midrule
    Training crop size & $512^2$ & $512^2$ & $1024^2$ \\
    Training images & $14{,}119$ & $14{,}119$ & $4{,}807$ \\
    Batch size (per GPU) & $8$ ($4$) & $32$ ($2$) & $8$ ($1$) \\
    Gradient accumulation & $1$ & $8$ & $4$ \\
    Steps & $350$k & $45$k & $10$k \\
    Initial learning rate & $10^{-4}$ & $5 \times 10^{-5}$ & $2 \times 10^{-5}$ \\
    Final learning rate & $5 \times 10^{-5}$ & $10^{-6}$ & $10^{-6}$ \\
    Gradient clipping & $1.0$ & $0.1$ & $1.0$ \\
    Perceptual loss $\mathcal{L}_p$ & LPIPS & DISTS & DISTS \\
    Rate factor $c$ & $1$ & $3$ & $3$ \\
    $\beta$ & $0$ & $[0, 5.12]$ & $[0, 5.12]$ \\
    \bottomrule
  \end{tabular}
\end{table}

\paragraph{Training Dataset}
We train on the $800$ training images of DIV2K~\cite{agustsson2017ntire} and the $2{,}650$ images of Flickr2K~\cite{lim2017enhanced}, which together are known as DF2K. We also use the $585$ training and $41$ validation images of CLIC2020 Professional~\cite{toderici2020clic}, the $60$ test images of CLIC2021, and the first $10{,}000$ training images of LSDIR~\cite{li2023lsdir}. We discard images whose shorter side is below $512$ pixels. This removes $17$ of the CLIC images and leaves $14{,}119$ images for stages~1 and~2. Stage~3 uses only the $4{,}807$ of these images whose shorter side is at least $1024$ pixels. Of these, $3{,}421$ come from DF2K, $562$ from CLIC and $824$ from LSDIR.

\paragraph{Optimization}
All stages train in fp32 on two NVIDIA A40 GPUs with AdamW~\cite{loshchilov2019decoupled}, using PyTorch's default momentum parameters and a weight decay of $0.01$. Each stage starts from the weights of the previous one with a fresh optimizer. The learning rate starts at $10^{-4}$, $5 \times 10^{-5}$ and $2 \times 10^{-5}$ in stages~1, 2 and~3, and step decay lowers it over each stage to $5 \times 10^{-5}$, $10^{-6}$ and $10^{-6}$, respectively. Training crops are taken at random positions and flipped horizontally and vertically at random. 

\paragraph{Loss Weights}
We set $\gamma_{\mathrm{dist}} = 2$, $\gamma_s = 0.1$, $\gamma_p = 1$, $\gamma_{\mathrm{adv}} = 0.05$ and, for \modelname{}-FD, $\gamma_{\mathrm{FD}} = 0.25$. The squared error is averaged over the pixels of images scaled to $[-1, 1]$. In stages~2 and~3, the rate term $\lambda\,\mathcal{R}$ of \cref{eq:loss_s2} is multiplied by a constant factor $c = 3$ to offset the rate increase caused by the realism losses, while the codec is still conditioned on the unscaled $\lambda$. During training, $\lambda$ is sampled for each image and $\beta$ for each micro-batch, as described below.

\paragraph{Sampling of $\lambda$ and $\beta$}
$\lambda$ is drawn for each image log-uniformly from $[0.05, 32]$ except for the initial steps of stage~1, where the range is $[0.05, 2]$ for the first $5$k steps and $[0.05, 8]$ for the next $5$k. Unlike $\lambda$, $\beta$ is shared by all images of a micro-batch. We split $[0, 5.12]$ into four buckets of width $1.28$ and assign them to the micro-batches in turn, across GPUs and accumulation steps. Each micro-batch then draws $\beta$ uniformly from its bucket, so every optimizer step covers the whole range of $\beta$.

\paragraph{Adversarial Loss}
Each head of the discriminator $\mathbf{D}$~(\cref{sec:supp_discriminator}) is trained with the binary cross-entropy corresponding to \cref{eq:loss_adv}. For the generator, we use the non-saturating loss. Instead of minimizing $\log(1 - \mathbf{D}_{\lambda,\beta}(\hat{x}))$ as in \cref{eq:loss_adv}, the generator minimizes
\begin{equation}
\mathcal{L}_{\mathrm{G}} = -\mathbb{E}\big[\log \mathbf{D}_{\lambda,\beta}(\hat{x})\big].
\label{eq:loss_gen}
\end{equation}
This rewards the generator when $\mathbf{D}$ classifies its reconstructions as real. Both losses are summed over the three heads. We update $\mathbf{D}$ once per generator step with AdamW at a constant learning rate of $2 \times 10^{-5}$. It is initialized at the start of stage~2 and carried over to stage~3.

\paragraph{FD-Loss}
For each of its three backbones, Inception-v3, SigLIP~2 and MAE, \modelname{}-FD keeps one queue of $N = 40{,}000$ features per $\beta$ bucket, so that reconstructions at different $\beta$ are not pooled into one distribution. FD is computed only in the three upper buckets, $\beta \geq 1.28$, since reconstructions at the distortion end are not meant to match the real distribution. The FD-Loss is only computed once a bucket's queue is full. Each training image crop is cut into non-overlapping $256 \times 256$ patches, $4$ patches per training image in stage~2 and $16$ in stage~3, which enter the queue as separate features.

\begin{figure*}[t]
  \centering
  \includegraphics[width=0.90\linewidth]{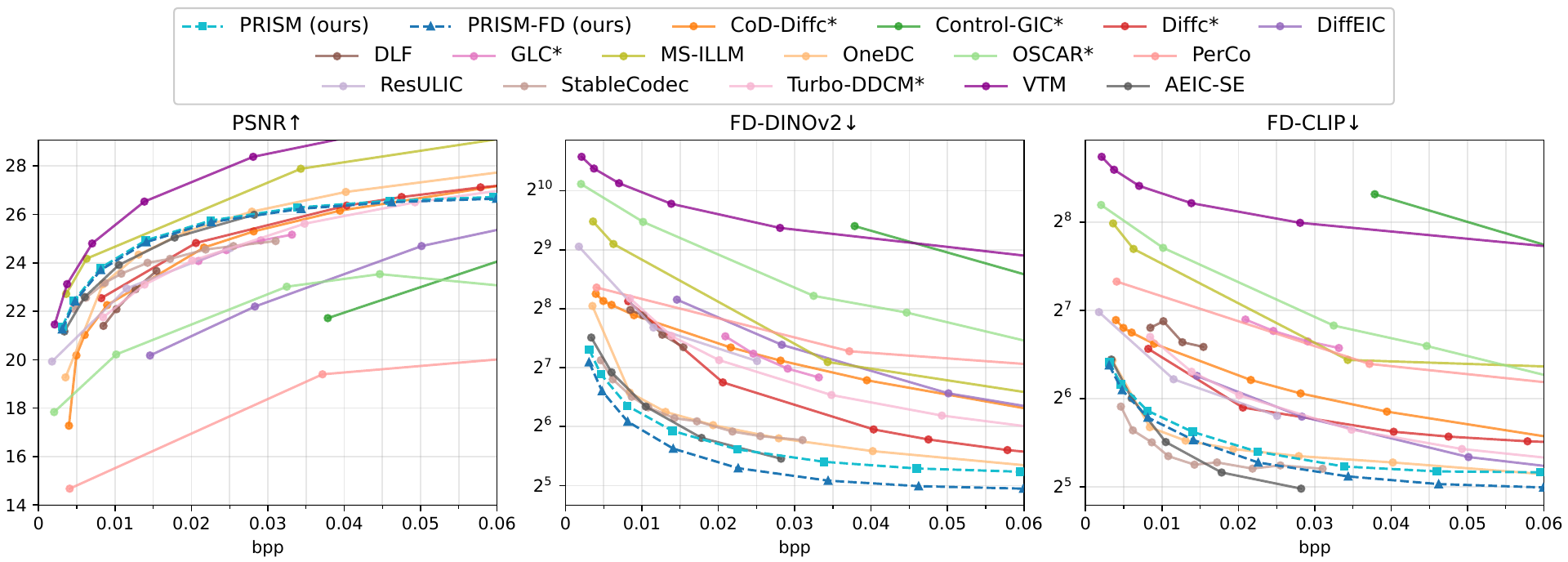}
  \caption{\textbf{Rate--distortion--perception comparison on CLIC2020 in PSNR, FD-DINOv2 and FD-CLIP}, which complete the CLIC2020 panels of \cref{fig:rd}.
    $\uparrow$/$\downarrow$ represents higher/lower is better.
    Codecs marked with * are variable-rate; the others, except VTM, train a separate model for each rate.
    \modelname{} and \modelname{}-FD are shown at $\beta = 5.12$.}
  \label{fig:supp_rd_clic}
\end{figure*}

\begin{table*}[t]
  \centering
  \caption{\textbf{BD-rate (\%, lower is better) of all codecs relative to CoD-Diffc*} on DIV2K, Kodak and CLIC2020, computed over the quality range in which each codec overlaps CoD-Diffc* (\cref{sec:supp_eval_metrics}). The best value in each column is in bold and the second best is underlined. FD-D and FD-C denote FD-DINOv2 and FD-CLIP. Codecs marked with * are variable-rate. \modelname{} and \modelname{}-FD are evaluated at $\beta = 5.12$ on DIV2K and CLIC2020 and at $\beta = 1.28$ on Kodak. PerCo and ResULIC have only three rate points, too few for BD-rate. The FID BD-rate of VTM on the CLIC2020 dataset is not computed as its quality range does not overlap with the reference.}
  \vspace{-0.2cm}
  \label{tab:supp_bd_rates}
  \footnotesize
  \setlength{\tabcolsep}{3pt}
  \renewcommand{\arraystretch}{1.1}
  \begin{tabular}{@{}lccccccccccccccc@{}}
    \toprule
    & \multicolumn{6}{c}{DIV2K} & \multicolumn{3}{c}{Kodak} & \multicolumn{6}{c}{CLIC2020} \\
    \cmidrule(lr){2-7} \cmidrule(lr){8-10} \cmidrule(l){11-16}
    Codec & PSNR & LPIPS & DISTS & FID & FD-D & FD-C & PSNR & LPIPS & DISTS & PSNR & LPIPS & DISTS & FID & FD-D & FD-C \\
    \midrule
    CoD-Diffc*~\cite{jia2026cod} & \multicolumn{15}{c}{reference} \\
    Control-GIC*~\cite{li2025once} & +150 & +135 & +442 & +381 & +356 & +855 & +138 & +119 & +411 & +183 & +118 & +435 & +408 & +379 & +1{,}787 \\
    Diffc*~\cite{vonderfecht2025lossy} & $-$4.0 & +2.2 & +48.8 & +21.6 & $-$10.4 & +7.2 & $-$12.6 & $-$6.6 & $-$3.2 & $-$6.4 & +0.9 & +46.0 & +32.5 & $-$37.4 & $-$27.2 \\
    GLC*~\cite{jia2024generative} & +18.7 & $-$26.6 & $-$3.5 & $-$27.3 & +2.9 & +53.3 & +32.0 & $-$24.6 & $-$10.7 & +22.6 & $-$28.1 & $-$6.3 & $-$12.2 & $-$0.4 & +312 \\
    OSCAR*~\cite{guo2025oscar} & +110 & +146 & +198 & +145 & +165 & +122 & +85.3 & +114 & +101 & +146 & +192 & +252 & +200 & +170 & +148 \\
    Turbo-DDCM*~\cite{vaisman2026turbo} & +4.2 & +25.4 & +3.5 & $-$6.3 & $-$7.0 & $-$20.9 & +8.3 & +16.0 & $-$9.3 & +11.7 & +31.0 & +2.6 & $-$8.3 & $-$16.4 & $-$26.1 \\
    \midrule
    AEIC-SE~\cite{zhang2026ultra} & $-$27.7 & $-$47.3 & $-$52.7 & $-$72.3 & $-$70.9 & \underline{$-$80.4} &$-$16.9 & $-$36.4 & $-$44.0 & $-$35.2 & $-$50.3 & $-$53.1 & $-$75.8 & $-$81.7 & $-$81.7 \\
    DiffEIC~\cite{li2024toward} & +133 & +79.4 & +173 & +122 & +54.5 & +23.7 & +123 & +65.4 & +89.4 & +189 & +93.6 & +206 & +235 & +27.3 & $-$35.8 \\
    DLF~\cite{xue2025dlf} & +24.1 & $-$22.1 & +13.9 & $-$31.8 & +43.9 & +1.6 & +11.3 & $-$25.6 & $-$7.6 & +18.2 & $-$27.9 & $-$3.0 & $-$32.9 & $-$7.8 & +61.8 \\
    MS-ILLM~\cite{muckley2023improving} & \underline{$-$61.8} & $-$2.7 & +175 & +144 & +90.5 & +194 & \underline{$-$59.5} & +2.7 & +203 & \underline{$-$61.3} &$-$16.1 & +104 & +84.7 & +46.3 & +213 \\
    OneDC~\cite{xue2025one} & $-$23.8 & \textbf{\boldmath$-$53.4} & $-$24.3 & $-$39.3 & $-$65.6 & $-$65.3 & $-$17.6 & \textbf{\boldmath$-$48.8} & $-$24.8 & $-$26.5 & \textbf{\boldmath$-$58.3} & $-$34.5 & $-$35.0 & $-$78.0 & \underline{$-$83.1} \\
    PerCo~\cite{careil2024towards} & \multicolumn{15}{c}{only three rate points} \\
    ResULIC~\cite{ke2025ultra} & \multicolumn{15}{c}{only three rate points} \\
    StableCodec~\cite{zhang2025stablecodec} & $-$21.1 & $-$31.0 & $-$30.3 & $-$63.9 & $-$66.8 & $-$78.0 & $-$30.9 & $-$31.9 & $-$29.5 & $-$19.6 & $-$35.1 & $-$27.6 & $-$65.8 & $-$81.0 & \textbf{\boldmath$-$86.7} \\
    \midrule
    VTM~\cite{bross2021overview} & \textbf{\boldmath$-$67.9} & +741 & +1{,}821 & +4{,}939 & +979 & +10{,}498 & \textbf{\boldmath$-$65.7} & +766 & +1{,}676 & \textbf{\boldmath$-$69.6} & +1{,}076 & +3{,}031 & -- & +1{,}615 & +9{,}406 \\
    \midrule
    \textbf{\modelname{}}* (ours) & $-$35.2 & \underline{$-$48.3} & \textbf{\boldmath$-$60.4} & \textbf{\boldmath$-$76.2} & \underline{$-$75.4} & $-$79.9 & $-$29.3 & \underline{$-$41.1} & \textbf{\boldmath$-$49.3} & $-$41.9 & \underline{$-$51.5} & \textbf{\boldmath$-$60.5} & \underline{$-$78.7} & \underline{$-$85.3} & $-$76.2 \\
    \textbf{\modelname{}-FD}* (ours) & $-$32.8 & $-$47.6 & \underline{$-$58.6} & \underline{$-$76.1} & \textbf{\boldmath$-$79.7} & \textbf{\boldmath$-$83.7} & $-$26.6 & $-$40.6 & \underline{$-$48.8} & $-$40.0 & $-$51.0 & \underline{$-$59.0} & \textbf{\boldmath$-$79.1} & \textbf{\boldmath$-$89.0} & $-$79.9 \\
    \bottomrule
  \end{tabular}
\end{table*}

\begin{table*}[t]
  \centering
  \caption{\textbf{Computational complexity of all codecs on Kodak and DIV2K.} We report latency in seconds and MACs in units of $10^{12}$ (TMACs), averaged over the $10$ benchmark images of each dataset and the rate points of each codec's rate--distortion curve. \Cref{sec:supp_complexity} describes the protocol. Codecs marked with * are variable-rate.
    \emph{Rates}: the rate points one model supports; for a fixed-rate codec, this is the number of released models.
    \emph{Params}: Number of network parameters in billions, summed over all models a codec needs to cover its rate range; for PerCo and ResULIC, the parameters and the encoding cost include their BLIP-2 captioner.
    \emph{Steps}: Number of denoising steps needed on average per decoding pass.
    \modelname{}-FD has the same cost as \modelname{}.
    $^\dagger$Over $9$ of the $10$ images; the tenth overflows in fp16.}
    \vspace{-0.2cm}
  \label{tab:complexity_full}
  \small
  \setlength{\tabcolsep}{4pt}
  \begin{tabular}{@{}lccccccccccc@{}}
    \toprule
    & & & & \multicolumn{4}{c}{Kodak ($768 \times 512$)} & \multicolumn{4}{c}{DIV2K ($2040 \times 1356$ to $2040 \times 1548$)} \\
    \cmidrule(lr){5-8} \cmidrule(l){9-12}
    & & & & \multicolumn{2}{c}{Latency (s)} & \multicolumn{2}{c}{TMACs} & \multicolumn{2}{c}{Latency (s)} & \multicolumn{2}{c}{TMACs} \\
    \cmidrule(lr){5-6} \cmidrule(lr){7-8} \cmidrule(lr){9-10} \cmidrule(l){11-12}
    Codec & Rates & Params (B) & Steps & Enc. & Dec. & Enc. & Dec. & Enc. & Dec. & Enc. & Dec. \\
    \midrule
    CoD-Diffc*~\cite{jia2026cod} & 160 & 0.94 & 96.6 & 9.87 & 9.94 & 78.3 & 103 & 120\rlap{$^\dagger$} & 132\rlap{$^\dagger$} & 1{,}231 & 1{,}608 \\
    Control-GIC*~\cite{li2025once} & $\infty$ & 0.13 & -- & 0.228 & 1.10 & 1.19 & 4.12 & \multicolumn{4}{c}{out of memory} \\
    Diffc*~\cite{vonderfecht2025lossy} & 161 & 1.07 & 107 & 8.28 & 6.26 & 64.8 & 71.2 & 136 & 125 & 1{,}195 & 1{,}302 \\
    GLC*~\cite{jia2024generative} & 4 & 0.10 & -- & 0.0867 & 0.134 & 0.475 & 0.817 & 0.647 & 1.05 & 3.65 & 6.27 \\
    OSCAR*~\cite{guo2025oscar} & 8 & 1.05 & 1 & 0.136 & 0.441 & 0.868 & 2.58 & 1.60 & 5.60 & 8.73 & 29.2 \\
    Turbo-DDCM*~\cite{vaisman2026turbo} & 16{,}384 & 0.95 & 30 & 1.50 & 1.53 & 20.3 & 21.3 & 21.4 & 21.7 & 369 & 377 \\
    \midrule
    AEIC-SE~\cite{zhang2026ultra} & 5 & 4.43 & 1 & 0.0282 & 0.246 & 0.0181 & 1.21 & 0.129 & 4.43 & 0.138 & 17.3 \\
    DiffEIC~\cite{li2024toward} & 5 & 6.83 & 50 & 0.264 & 8.56 & 0.920 & 35.4 & 2.24 & 166 & 9.03 & 654 \\
    DLF~\cite{xue2025dlf} & 4 & 4.96 & -- & 0.220 & 0.373 & 1.07 & 1.97 & 1.67 & 3.03 & 8.72 & 16.6 \\
    MS-ILLM~\cite{muckley2023improving} & 8 & 1.45 & -- & 0.0909 & 0.121 & 0.0385 & 0.266 & 0.519 & 0.758 & 0.295 & 2.04 \\
    OneDC~\cite{xue2025one} & 8 & 11.25 & 1 & 0.233 & 0.631 & 1.05 & 2.68 & 2.28 & 7.67 & 8.50 & 27.8 \\
    PerCo~\cite{careil2024towards} & 3 & 7.63 & 15 & 0.681 & 5.25 & 1.71 & 21.6 & 2.34 & 86.7 & 9.45 & 378 \\
    ResULIC~\cite{ke2025ultra} & 3 & 7.82 & 4 & 0.701 & 1.51 & 1.70 & 7.40 & 2.54 & 30.7 & 10.0 & 131 \\
    StableCodec~\cite{zhang2025stablecodec} & 9 & 9.59 & 1 & 0.219 & 0.471 & 1.03 & 2.61 & 2.17 & 6.04 & 10.6 & 31.5 \\
    \midrule
    \textbf{\modelname{}}* (ours) & 256 & 1.01 & 1 & 0.0648 & 0.281 & 0.123 & 1.30 & 0.376 & 3.97 & 0.934 & 18.0 \\
    \bottomrule
  \end{tabular}
\end{table*}

\section{Evaluation Metrics}
\label{sec:supp_eval_metrics}

We score the reconstructions of every codec, including the baselines, with the same implementation of each quality metric. Let $x$ be an image of $H \times W$ pixels with RGB values in $[0, 1]$ and $\hat{x}$ its reconstruction. The bitrate is the length of the bitstream in bits divided by $HW$.

\paragraph{Full-Reference Metrics}
PSNR is computed over the three RGB channels,
\begin{equation}
  \mathrm{PSNR}(x, \hat{x}) = 10 \log_{10} \frac{1}{\mathrm{MSE}(x, \hat{x})},
\end{equation}
where
\begin{equation}
    \mathrm{MSE}(x, \hat{x}) = \frac{1}{3HW}{\|x - \hat{x}\|_2^2}.
\end{equation}
LPIPS~\cite{zhang2018unreasonable} compares the features of a pre-trained AlexNet model. Let $\phi_l^{hw}(x) \in \mathbb{R}^{C_l}$ be the activations of layer $l$ at position $(h, w)$ of its $H_l \times W_l$ grid, normalized to unit length over the channels. Then
\begin{equation}
  \mathrm{LPIPS}(x, \hat{x}) = \sum_{l} \frac{1}{H_l W_l} \sum_{h, w} \bigl\| w_l \odot \Delta\phi_l^{hw} \bigr\|_2^2,
\end{equation}
where $\Delta\phi_l^{hw} = \phi_l^{hw}(x) - \phi_l^{hw}(\hat{x})$ and $w_l \in \mathbb{R}^{C_l}$ are learned channel weights. DISTS~\cite{ding2020image} compares the structure and texture of the image itself and of the outputs of five stages of a pre-trained VGG-16 model. For channel $j$ of stage $i$, let $m_{ij}$ and $\hat{m}_{ij}$ be the spatial means of the feature maps of $x$ and $\hat{x}$, $v_{ij}$ and $\hat{v}_{ij}$ their spatial variances, and $c_{ij}$ their covariance. Then
\begin{equation}
  \mathrm{DISTS}(x, \hat{x}) = 1 - \sum_{i, j} \bigl( a_{ij}\, \ell_{ij} + b_{ij}\, s_{ij} \bigr),
\end{equation}
with the texture and structure terms
\begin{equation}
  \ell_{ij} = \frac{2 m_{ij} \hat{m}_{ij} + \epsilon_1}{m_{ij}^2 + \hat{m}_{ij}^2 + \epsilon_1}, \qquad
  s_{ij} = \frac{2 c_{ij} + \epsilon_2}{v_{ij} + \hat{v}_{ij} + \epsilon_2},
\end{equation}
where the weights $a_{ij}, b_{ij} \geq 0$ are learned and sum to $1$, and $\epsilon_1$ and $\epsilon_2$ are small constants. The three metrics are computed on whole image pairs and averaged over the entire dataset.

\paragraph{Distribution Metrics}
FID~\cite{heusel2017gans}, FD-DINOv2 and FD-CLIP are the FD of \cref{eq:supp_fd_common} between the features of original and reconstructed patches. They differ only in the backbone $\Phi$:
\begin{itemize}
  \item FID uses the $2048$-dimensional pooled features of Inception-v3~\cite{szegedy2016rethinking}, with each patch resized to $299 \times 299$.
  \item FD-DINOv2 uses the $1024$-dimensional class token of DINOv2 ViT-L/14~\cite{oquab2024dinov2}, with each patch resized to $224 \times 224$.
  \item FD-CLIP uses the $1024$-dimensional class token of the image encoder of OpenAI's CLIP ViT-L/14~\cite{radford2021learning}, taken before its final projection, with each patch resized to $224 \times 224$.
\end{itemize}
Following \citet{mentzer2020high}, each image is split into a grid of non-overlapping $256 \times 256$ patches, and a second grid, shifted by $128$ pixels in both directions, adds further patches. This gives $6{,}573$ patches for DIV2K and $28{,}650$ for CLIC2020. Kodak gives only $192$ patches, too few to estimate covariances of $1024$ or $2048$ dimensions, so we report only the full-reference metrics on Kodak.

\paragraph{BD-Rate}
For each codec and metric, we fit the log-rate $r = \log_{10} \mathrm{bpp}$ as a cubic polynomial $r(q)$ of the quality $q$, by least squares over all rate points of the codec~\cite{bjontegaard2001calculation}, as in QARV~\cite{duan2023qarv}. The BD-rate of a test codec relative to a reference codec is the average difference of the two fits over the quality range $[q_{\mathrm{lo}}, q_{\mathrm{hi}}]$ in which both curves overlap,
\begin{align}
  \Delta r &= \frac{1}{q_{\mathrm{hi}} - q_{\mathrm{lo}}} \int_{q_{\mathrm{lo}}}^{q_{\mathrm{hi}}} \bigl( r_{\mathrm{test}}(q) - r_{\mathrm{ref}}(q) \bigr)\, \mathrm{d}q, \\
  \text{BD-rate} &= \bigl(10^{\Delta r} - 1\bigr) \times 100\%,
\end{align}
so a negative BD-rate means that the test codec needs fewer bits for the same quality. We do not compute the BD-rate for codecs with fewer than four rate points. For the perceptual metrics, we transform $q$ to a logarithmic form in which higher is better: $-10 \log_{10} \mathrm{LPIPS}$, $-10 \log_{10} \mathrm{DISTS}$, $-10 \log_{10} (\mathrm{FID} / 100)$, and $-10 \log_{10} (\mathrm{FD} / 1000)$ for FD-DINOv2 and FD-CLIP. The constants only shift the axis and do not change the BD-rate. All BD-rates in the paper are computed in this way.

\section{Rate--Distortion and Rate--Perception Performance}
\label{sec:supp_quantitative_comparison}

This section completes the comparison of \cref{sec:exp_baselines} with the CLIC2020 curves that \cref{fig:rd} leaves out and with the BD-rates of all codecs on all three datasets.

\paragraph{CLIC2020 Curves}
\Cref{fig:supp_rd_clic} shows the three metrics on CLIC2020 that \cref{fig:rd} leaves out. In PSNR, \modelname{} is behind only VTM and MS-ILLM, whose LPIPS, DISTS and FID are far worse (\cref{fig:rd}). In FD-DINOv2, \modelname{}-FD is the best codec, followed by \modelname{}. In FD-CLIP, StableCodec, OneDC and AEIC-SE are ahead of both \modelname{} models, and the FD-Loss closes part of the gap.

\paragraph{BD-Rates}
\Cref{tab:supp_bd_rates} lists the BD-rates of all codecs in all plotted metrics. We use CoD-Diffc as the reference because its quality range overlaps that of every other codec in every metric, with the only exception of VTM in FID on CLIC2020. It is a single variable-rate model with ten rate points from $0.004$ to $0.11$--$0.13$\,bpp. Each BD-rate is computed only over the range where a codec overlaps with CoD-Diffc. The difference between two rows therefore only approximates a direct comparison of the two codecs. For this reason, the BD-rates of \modelname{} relative to AEIC-SE in \cref{sec:exp_baselines} are computed directly between the two.

\modelname{} has the lowest BD-rate in DISTS on all three datasets. In FID, on DIV2K and CLIC2020, \modelname{} and \modelname{}-FD have the lowest BD-rates. \modelname{}-FD leads in FD-DINOv2 on both of these datasets and in FD-CLIP on DIV2K, whereas in FD-CLIP on CLIC2020, StableCodec, OneDC and AEIC-SE are ahead. In LPIPS, \modelname{} is second only to OneDC on all three datasets. In PSNR, it is behind only VTM and MS-ILLM on DIV2K and CLIC2020. Among the variable-rate codecs, \modelname{} is the best in every metric on every dataset.

\section{Computational Complexity}
\label{sec:supp_complexity}

\Cref{tab:complexity_full} reports the computational complexity of \modelname{} and of every baseline in our rate--distortion comparison, on both Kodak and DIV2K. VTM, which runs on the CPU, is left out.

\paragraph{Benchmark Images}
Every codec codes the same $10$ randomly chosen images of each dataset. The Kodak images (kodim02, 06, 09, 10, 13, 14, 16, 17, 19 and 24) all have $768 \times 512$ pixels. The DIV2K validation images (0813, 0818, 0828, 0833, 0837, 0865, 0869, 0875, 0880 and 0897) range from $2040 \times 1356$ to $2040 \times 1548$ pixels.

\paragraph{Protocol}
We run every codec on the same NVIDIA Quadro RTX 6000 GPU (24\,GB), with no other process on the GPU. Each codec codes every image in a single pass, without tiling, at every rate point of its rate--distortion curve in \cref{fig:rd}, and we report the mean over the images and these rate points. Latency is measured as the CUDA-synchronized wall-clock time after warm-up. Encoding runs from the image tensor on the GPU to the bitstream, including entropy coding, and decoding from the bitstream to the reconstructed image tensor. MACs are the multiply--accumulate operations of convolutions, matrix products and attention, counted over every PyTorch operation a codec executes. For \modelname{}, whose selective scan runs in a custom CUDA kernel, we add the MACs of the scan analytically, following the count of VMamba~\cite{liu2024vmamba}. \emph{Params} counts the network parameters of a codec summed over all models the codec needs to cover its rate range. PerCo and ResULIC caption the input image with BLIP-2~\cite{li2023blip}, whose $3.75$B parameters we count once and include, with the cost of captioning, in their parameters and encoding cost. \emph{Rates} is the number of rate points one model supports; for a fixed-rate codec, it is the number of released models. \emph{Steps} is the number of denoising steps per decoding pass, averaged over the rate points. Diffc and Turbo-DDCM run in fp16, CoD-Diffc with bf16 weights and fp16 autocast for its diffusion transformer, OneDC under bf16 autocast, and all other codecs, including \modelname{} and AEIC-SE, in fp32. \modelname{}-FD has the same architecture as \modelname{} and therefore the same cost.

\paragraph{Results}
\modelname{} covers $256$ rate points with a single model of $1.01$B parameters, about as many as the other variable-rate diffusion codecs ($0.94$--$1.07$B), whereas the fixed-rate diffusion codecs need $4.4$--$11.3$B parameters for their $3$--$9$ released rates. The $256$ rate points of \modelname{} are set by a $1$-byte index into a log-spaced grid of $\lambda$, which is part of the bitstream and included in all reported bitrates. Since $\lambda$ is sampled continuously in training, a $2$-byte index would let the same model cover $65{,}536$ rate points, at an extra cost of at most $2.0 \times 10^{-5}$\,bpp on the $768 \times 512$ Kodak images and below $5 \times 10^{-6}$\,bpp on DIV2K. Since it decodes with a single denoiser step, \modelname{} decodes a DIV2K image in $3.97$\,s, over $30\times$ faster than the multi-step variable-rate codecs Diffc and CoD-Diffc. On Kodak, AEIC-SE is the only diffusion-based codec that decodes faster ($0.25$\,s \vs $0.28$\,s). The codecs without a denoiser, MS-ILLM, GLC and DLF, decode faster still ($0.76$--$3.03$\,s on DIV2K). The encoder of \modelname{}, which includes the spatial context model, is the second fastest after that of AEIC-SE on both datasets.

\section{Additional Visual Results}
\label{sec:supp_qualitative_comparison}

\begin{figure*}[p]
  \centering
  \includegraphics[width=\linewidth,height=0.9\textheight,keepaspectratio]{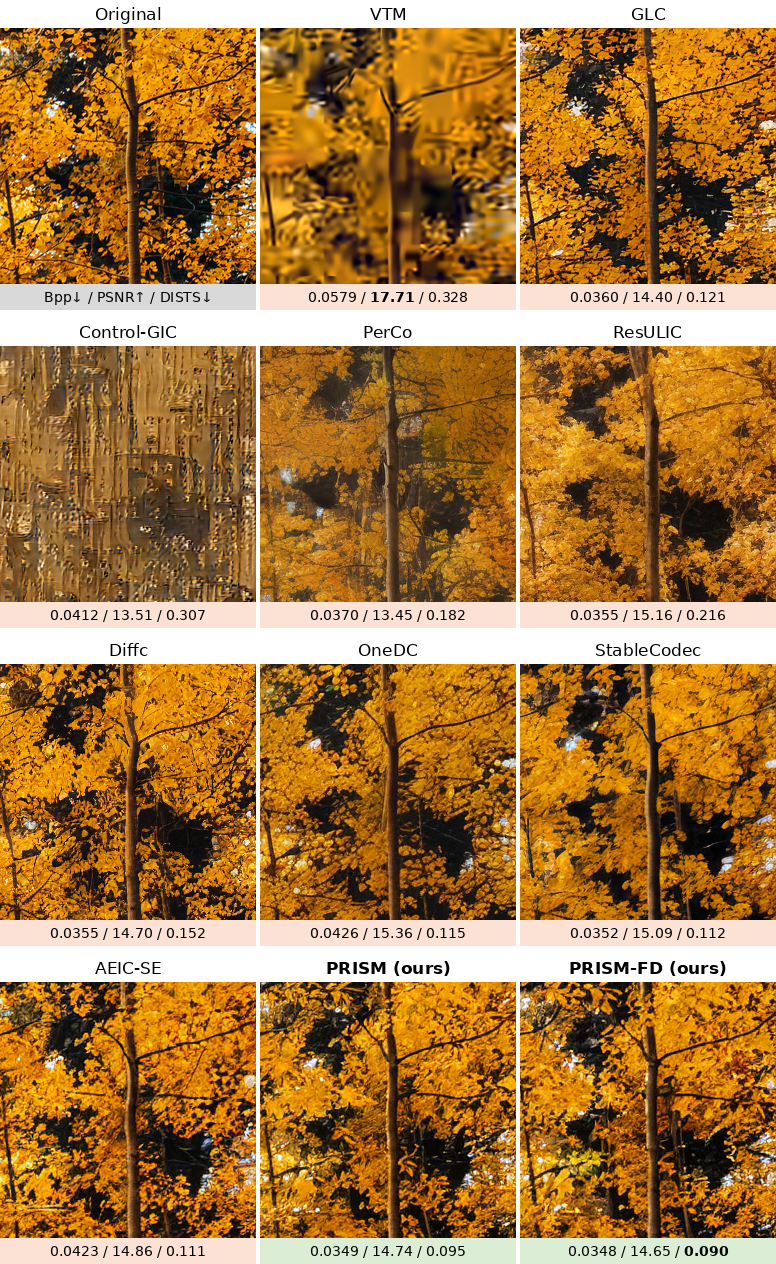}
  \caption{\textbf{Visual comparison on CLIC2020 image 0384}, coded by \modelname{} at $0.0349$\,bpp. Below each $256 \times 256$ crop are the bpp, PSNR and DISTS of the whole image, with the best PSNR and DISTS in bold. \modelname{} and \modelname{}-FD are shown at $\beta = 5.12$, and each baseline at its lowest rate point with at least as many bits as both (\cref{sec:supp_qualitative_comparison}). Both \modelname{} models keep the leaves and the dark gaps between them, which VTM blurs and Control-GIC replaces with a stripe pattern.}
  \label{fig:supp_qual_clic_0384}
\end{figure*}

\begin{figure*}[p]
  \centering
  \includegraphics[width=\linewidth,height=0.9\textheight,keepaspectratio]{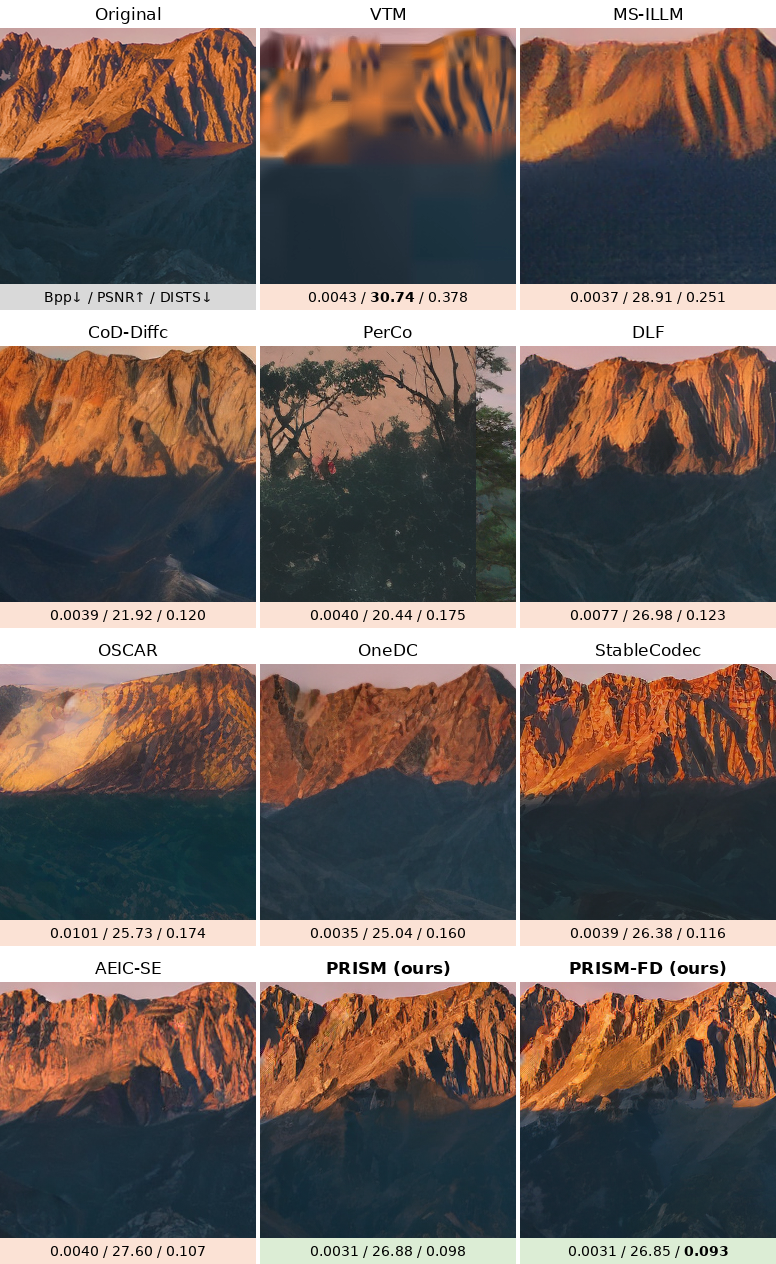}
  \caption{\textbf{Visual comparison on CLIC2020 image 0315}, coded by \modelname{} at $0.0031$\,bpp. The layout is as in \cref{fig:supp_qual_clic_0384}. \modelname{} keeps the sunlit ridges and the gullies between them, which VTM and MS-ILLM blur, OSCAR covers with haze and PerCo replaces with trees.}
  \label{fig:supp_qual_clic_0315}
\end{figure*}

\begin{figure*}[p]
  \centering
  \includegraphics[width=\linewidth,height=0.9\textheight,keepaspectratio]{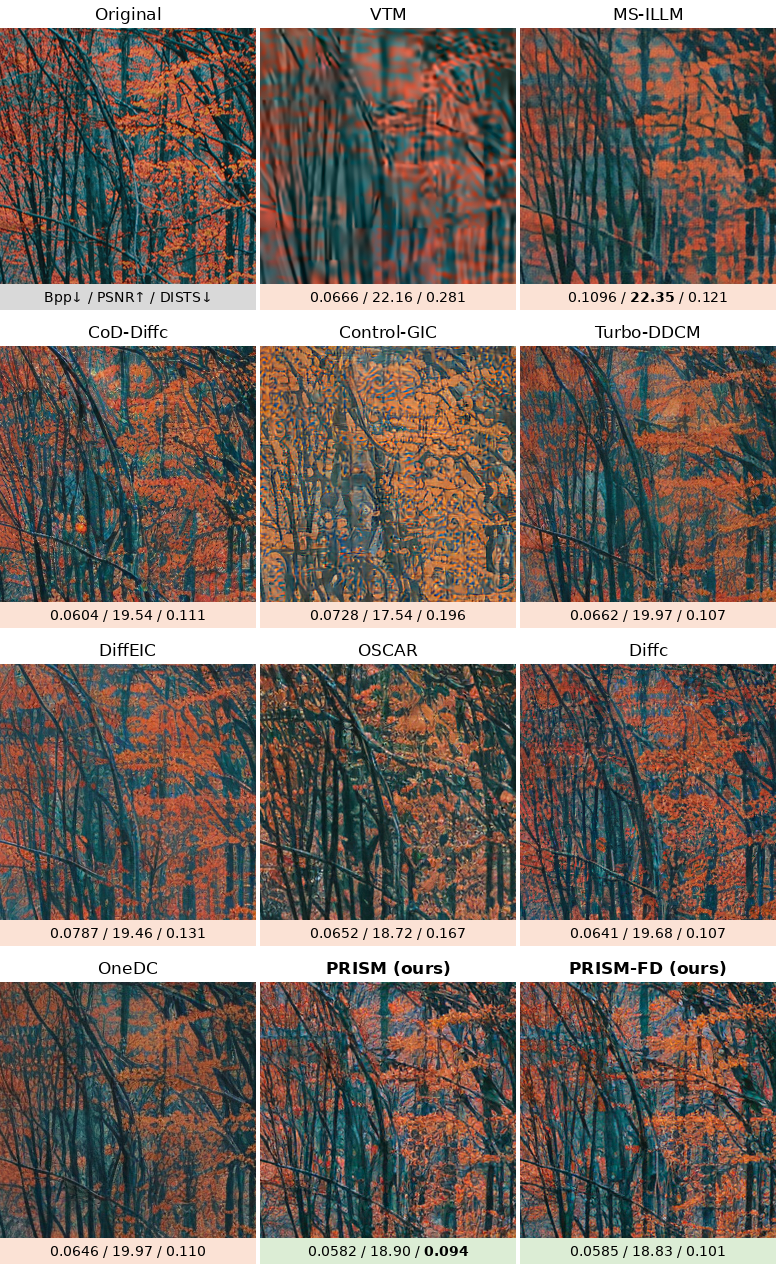}
  \caption{\textbf{Visual comparison on DIV2K image 0856}, coded by \modelname{} at $0.0582$\,bpp. The layout is as in \cref{fig:supp_qual_clic_0384}. \modelname{} keeps the thin trunks and the clusters of orange leaves, which VTM blurs and Control-GIC turns into a blocky pattern.}
  \label{fig:supp_qual_div2k_0856}
\end{figure*}

\begin{figure*}[p]
  \centering
  \includegraphics[width=\linewidth,height=0.9\textheight,keepaspectratio]{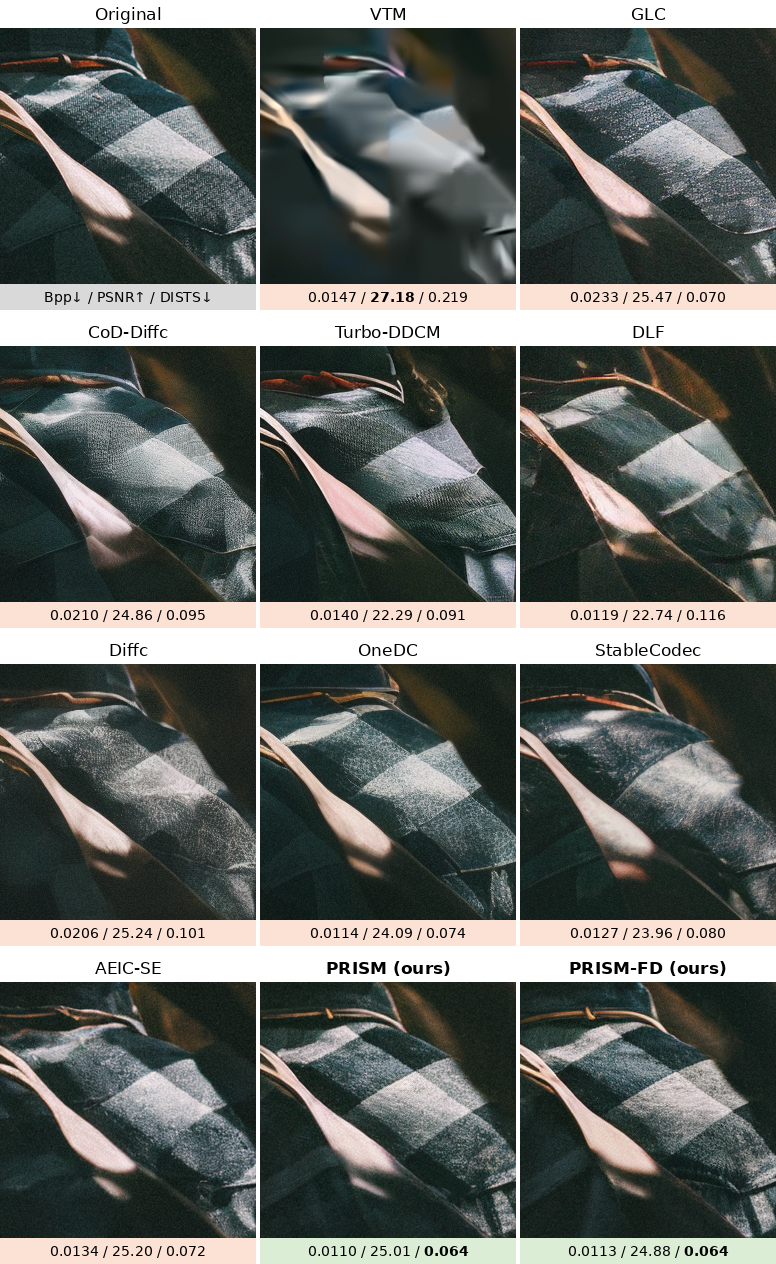}
  \caption{\textbf{Visual comparison on DIV2K image 0893}, coded by \modelname{} at $0.0110$\,bpp. The layout is as in \cref{fig:supp_qual_clic_0384}. Both \modelname{} models keep the checks and the weave of the fabric, whereas DLF and StableCodec blur the weave.}
  \label{fig:supp_qual_div2k_0893}
\end{figure*}

\begin{figure*}[p]
  \centering
  \includegraphics[width=\linewidth,height=0.9\textheight,keepaspectratio]{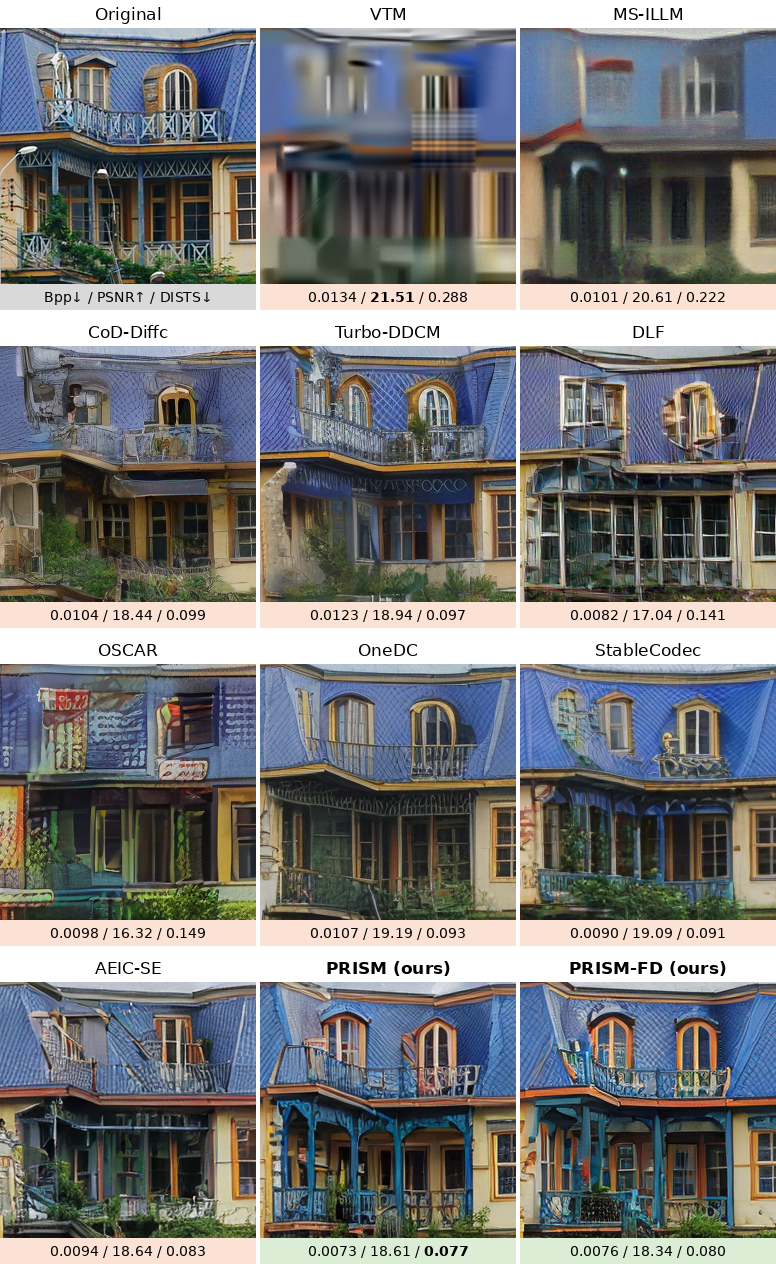}
  \caption{\textbf{Visual comparison on DIV2K image 0836}, coded by \modelname{} at $0.0073$\,bpp. The layout is as in \cref{fig:supp_qual_clic_0384}. \modelname{} keeps the two arched dormer windows, the balcony railing and the posts of the veranda, which VTM and MS-ILLM blur and OSCAR and DLF distort. AEIC-SE does not retain the arch shape of the windows.}
  \label{fig:supp_qual_div2k_0836}
\end{figure*}

\begin{figure*}[p]
  \centering
  \includegraphics[width=\linewidth,height=0.9\textheight,keepaspectratio]{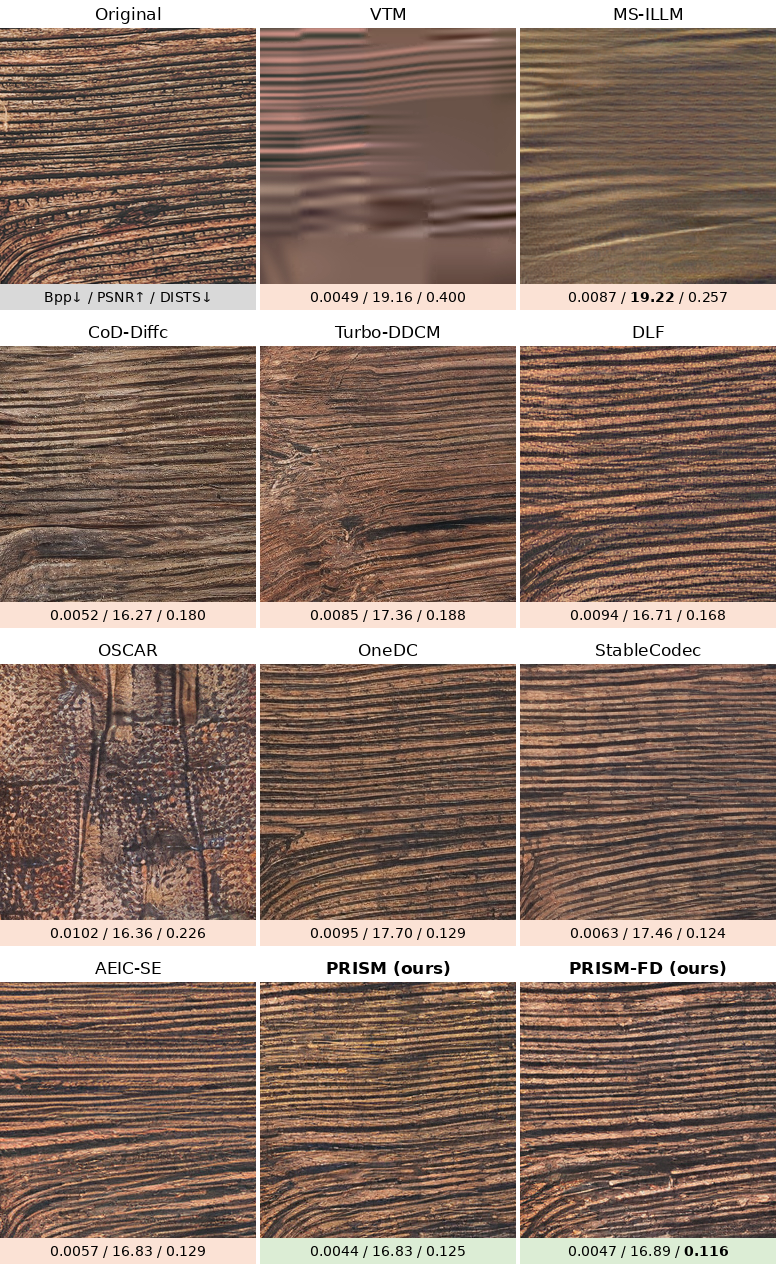}
  \caption{\textbf{Visual comparison on DIV2K image 0885}, coded by \modelname{} at $0.0044$\,bpp. The layout is as in \cref{fig:supp_qual_clic_0384}. Both \modelname{} models keep the fine grain of the wood, which VTM and MS-ILLM blur and OSCAR replaces with a cracked pattern.}
  \label{fig:supp_qual_div2k_0885}
\end{figure*}

\begin{figure*}[p]
  \centering
  \includegraphics[width=\linewidth,height=0.9\textheight,keepaspectratio]{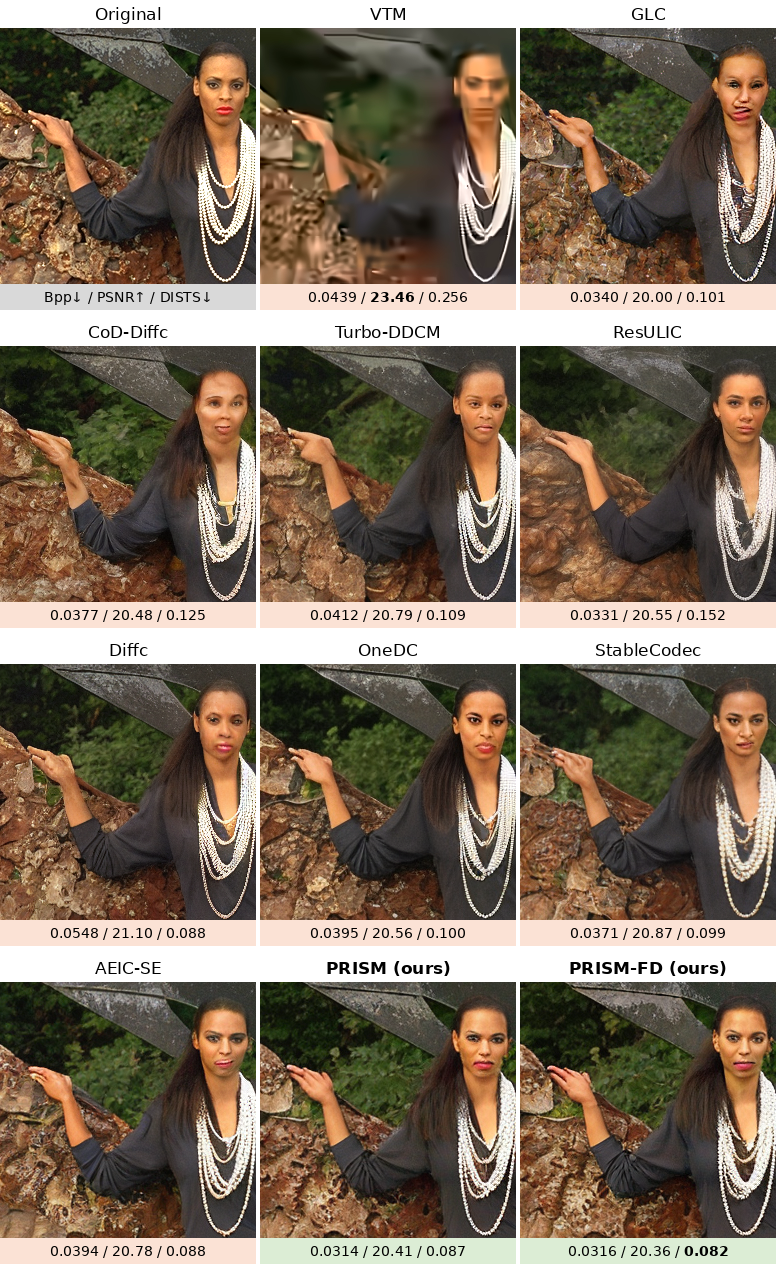}
  \caption{\textbf{Visual comparison on Kodak image kodim18}, coded by \modelname{} at $0.0314$\,bpp. The layout is as in \cref{fig:supp_qual_clic_0384}. Both \modelname{} models show the least amount of distortion on the face. VTM and ResULIC smooth the rock texture.}
  \label{fig:supp_qual_kodim18}
\end{figure*}

\begin{figure*}[p]
  \centering
  \includegraphics[width=\linewidth,height=0.9\textheight,keepaspectratio]{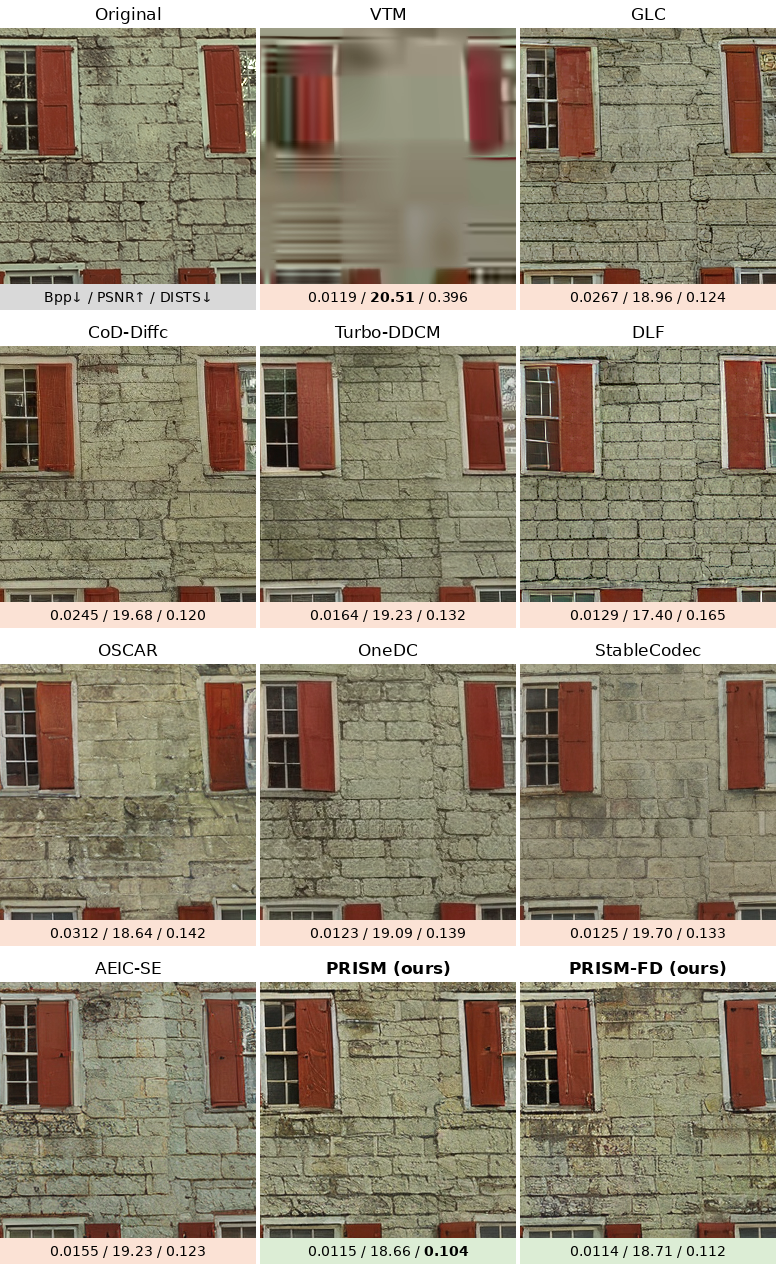}
  \caption{\textbf{Visual comparison on Kodak image kodim01}, coded by \modelname{} at $0.0115$\,bpp. The layout is as in \cref{fig:supp_qual_clic_0384}. \modelname{} keeps the shapes of the stone blocks and their rough surface, which DLF redraws as a regular brick pattern and StableCodec smooths.}
  \label{fig:supp_qual_kodim01}
\end{figure*}

\begin{figure*}[p]
  \centering
  \includegraphics[width=\linewidth,height=0.9\textheight,keepaspectratio]{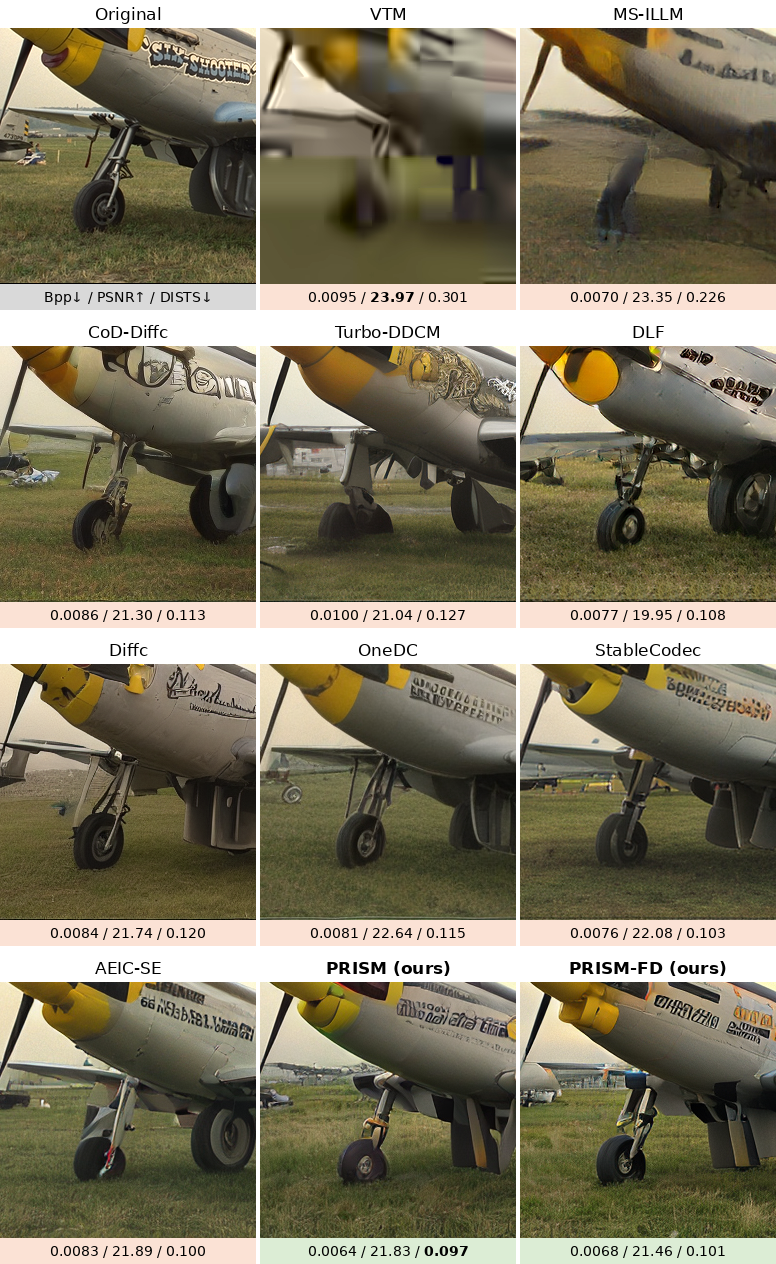}
  \caption{\textbf{Failure case: small text in Kodak image kodim20}, coded by \modelname{} at $0.0064$\,bpp. The layout is as in \cref{fig:supp_qual_clic_0384}. Although it has the best DISTS, \modelname{} replaces the nose art with illegible strokes and tints part of the yellow cowling green. VTM and MS-ILLM blur the text, and the other generative codecs also replace it with illegible glyphs.}
  \label{fig:supp_qual_kodim20}
\end{figure*}

\begin{figure*}[p]
  \centering
  \includegraphics[width=\linewidth,height=0.9\textheight,keepaspectratio]{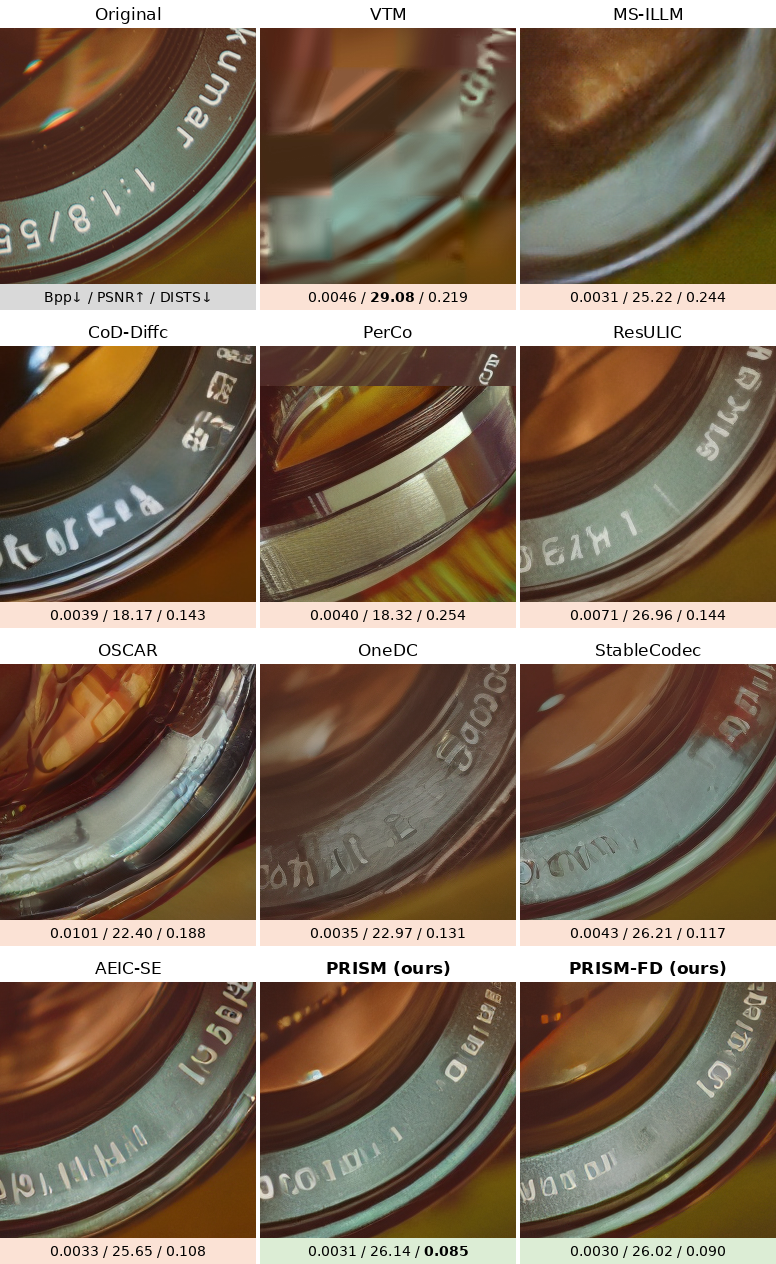}
  \caption{\textbf{Failure case: small text in CLIC2020 image 0293}, coded by \modelname{} at $0.0031$\,bpp. The layout is as in \cref{fig:supp_qual_clic_0384}. Although it has the best DISTS, \modelname{} replaces the engraved name, aperture and focal length on the lens barrel with illegible strokes. VTM and MS-ILLM blur the markings, PerCo and OSCAR lose them, and the other generative codecs also replace them with illegible glyphs.}
  \label{fig:supp_qual_clic_0293}
\end{figure*}

\Crefrange{fig:supp_qual_clic_0384}{fig:supp_qual_kodim01} compare \modelname{} and \modelname{}-FD with the baselines on eight further image crops of CLIC2020, DIV2K and Kodak, and \cref{fig:supp_qual_kodim20,fig:supp_qual_clic_0293} show two failure cases. Below each crop are the bpp, PSNR and DISTS of the whole image, and the best PSNR and DISTS of each collage are in bold. \modelname{} and \modelname{}-FD are shown at $\beta = 5.12$. Each baseline is shown at its lowest rate point that spends at least as many bits on the image as both \modelname{} models, so every baseline uses $1.01$--$3.3\times$ the bits of \modelname{}.

In every collage, either VTM or MS-ILLM has the best PSNR, but at the cost of blurred texture. \modelname{} uses fewer bits than every baseline, yet it keeps the structure of the scene, such as branches, ridges and window frames, and synthesizes texture such as leaves, fabric and stone. Unlike StableCodec in \cref{fig:supp_qual_div2k_0893,fig:supp_qual_kodim01}, \modelname{} generally does not smooth out texture. Nor does it generate entirely new content, as PerCo does in \cref{fig:supp_qual_clic_0315}.

\Cref{fig:supp_qual_kodim20,fig:supp_qual_clic_0293} show two images with small text: the nose art of the aircraft in kodim20, and the engraved name, aperture and focal length on the lens barrel in CLIC2020 image 0293. At $0.0064$ and $0.0031$\,bpp, \modelname{} replaces the letters with strokes that resemble text but cannot be read. No codec reconstructs the text at these rates: VTM and MS-ILLM blur it, and the other generative codecs either draw illegible glyphs as well or lose the text altogether.

\end{document}